\documentstyle[11pt,epsfig]{article}
[1999/03/29 PSNFSS v.7.2
Times font as default roman
: S Rahtz]

\newcommand{\be}{\begin{equation}}
\newcommand{\ee}{\end{equation}}
\newcommand{\ba}{\begin{eqnarray}}
\newcommand{\ea}{\end{eqnarray}}

\begin{document}

\title{Long-range Static Directional Stress Transfer \\in a Cracked, Nonlinear Elastic 
Crust}

\author{G. Ouillon$^{1,2}$ and D. Sornette$^{1,2}$\\
$^1$ Laboratoire de Physique de la
Mati\`{e}re Condens\'{e}e\\  CNRS UMR 6622 and Universit\'{e} de
Nice-Sophia Antipolis\\ Parc Valrose, 06108 Nice, France\\
$^2$ Department of Earth and Space Sciences\\ and Institute of
Geophysics and Planetary Physics\\ University of California, Los Angeles,
California, USA\\ e-mails: ouillon@aol.com and sornette@naxos.unice.fr}

\maketitle

\pagenumbering{arabic}

\begin{abstract}

Seeing the Earth crust as crisscrossed by faults filled with fluid
at close to lithostatic pressures, we develop
a model in which its elastic modulii are different in 
net tension versus compression. In constrast with standard nonlinear
effects, this ``threshold nonlinearity'' is non-perturbative and
occurs for infinitesimal perturbations around the lithostatic pressure
taken as the reference. For a given earthquake source,
such nonlinear elasticity is shown to (i) rotate, widen or narrow the
different lobes of stress transfer,
(ii) to modify the $1/r^2$ 2D-decay of elastic stress Green functions
into the generalized power law $1/r^{\gamma}$ 
where $\gamma$ depends on the azimuth and on the amplitude of the modulii asymmetry.
Using reasonable estimates, this implies an enhancement of the range
of interaction between earthquakes by a factor up to
$5-10$ at distances of several tens of rupture length. 
This may explain certain long-range earthquake triggering
and hydrological anomalies in wells and suggest to revisit 
the standard stress transfer calculations which use linear elasticity.
We also show that the standard double-couple of forces representing an earthquake source
leads to an opening of the corresponding fault plane, which suggests 
a mechanism for the non-zero isotropic component of the seismic moment tensor
observed for some events.

\end{abstract}

\section{Introduction}

There are many evidences that faults and earthquakes
interact, as suggested by calculations of stress redistribution
\cite{Stein0},
elastodynamic propagation of ruptures using laboratory-based friction
law \cite{Dieterich1,Dieterich2}, simplified
models of multiple faults
\cite{Cowie1,Cowie2}, as well as general
constraints of kinematic and geometric compatibility of the deformations
\cite{Gabrielovetal}. 
Maybe the simplest mechanism for earthquake interaction involves
stress re-distribution, both static
\cite{Harriscc,Stein0}
and dynamical \cite{Gomberg1} associated with a given earthquake modeled
as a set of dislocations or cracks. In this simple mechanical view,
earthquakes cast stress shadows in lobes 
of stress unloading \cite{stershs,Harriscc} and
increase the probability of rupture in zones of stress increase \cite{Harrissimprea},
according to the laws of linear elasticity. 
These elastic stress transfer models are useful for their
conceptual simplicity and are increasingly used. 
Notwithstanding their extended use,
the calculations of stress transfer have large
uncertainties stemming from (i) the usually
poorly known geometry of the rupture surfaces, (ii) the unconstrained 
homogeneity and
amplitude of the stress drop and/or of the slip distribution on
the fault plane, (iii) the use of simplified models of the 
crust (3D semi-infinite,
or thin elastic plate, or plate coupled to a semi-infinite 
visco-elastic asthenosphere, etc.),
and (iv) the unknown direction and amplitude of the absolute
stress field that pre-existed before the event, including its possible spatial
inhomogeneity.

Such elastic
stress transfer models seem unable to account for 
a growing phenomenology of long-range earthquake interactions. For instance,
many large earthquakes have been
preceded by an increase in the number of intermediate sized events
over very broad areas \cite{kbmal,Bowetal}. The
relation between these intermediate sized events and the subsequent main
event has only recently been recognized on a large
scale because the precursory events occur
over such a large area that they do not fit prior definitions of foreshocks
\cite{JonesMolnar}. In particular, the $11$
earthquakes in California with magnitudes greater than $6.8$ in the last
century are associated with an increase of precursory intermediate magnitude
earthquakes measured in a running time window of five years \cite{Knopoff}. What is
strange about the result is that the precursory pattern occured with
distances of the order of $300$ to $500 ~km$ from the futur epicenter, {\it
i.e.} at distances up to ten times larger that the size of the futur earthquake rupture. 
Furthermore, the increased intermediate magnitude activity switched
off rapidly after a big earthquake in about half of the cases. This implies
that stress changes due to an earthquake of rupture dimension as small as
$35 ~km$ can influence the stress distribution to distances more than ten
times its size. These observations of
earthquake-earthquake interactions over long times
and large spatial separations have been strengthened by several other
works on different catalogs using a variety of techniques \cite{Bowetal,Bowmanetal}.
These results defy usual mechanical models of linear elasticity
and one proposed explanation is that seismic cycles
represent the approach to and retreat from a critical state of a fault network
\cite{Bowmanetal,Sorsammis2}. Within the critical earthquake concept,
the anomalous long-range interactions between 
earthquakes reflect the increasing stress-stress
correlation length upon the approach of the critical earthquake.
Another explanation involves 
dynamical stress triggering \cite{kilb} (see however \cite{Gomberg2}).
Additional seismic, geophysical, and
hydrogeological observations \cite{ROELOFFS} cannot be accounted for by
using models derived from the elastic stress transfer mechanism. 
In particular, standard poro-elastic models underestimate
grossly the observed amplitudes of
hydrogeological anomalous rises and drops in wells at large distances from
earthquakes.

\section{Mechanical Model of the Earth's Crust}

Here, we investigate the hypothesis, and its implications
for the above observations, that the crust is a nonlinear 
elastic medium characterized by an asymmetric 
response to compressive versus extensive perturbations around
the lithostatic stress. We call this a ``threshold nonlinearity.''
This nonlinearity stems from a mechanically-justified
argumentation based on the fact that the Earth's crust at seismogenic depth is
crisscrossed by joints, cracks
or faults at many different scales filled with drained fluid 
in contact with delocalized reservoirs at pressures
close to the lithostatic pressure. 
It has been argued that rock permeability and
thus microcracking adjusts itself, so that fluid pressure is always close to rock
pressure irrespective of the extend of hydration/dehydration 
\cite{com,Wal}. One possible mechanism for this 
involves a time-dependent process that relates fluid pressure, flow pathways and fluid
volumes \cite{Nurl}.

\subsection{Presence and role of fluids \label{jfgkal}}

Indeed, a lot of data collectively support the existence of significant fluid circulation 
to crustal depths of at least $10-15 km$. 
Much attention
has been devoted to the role of overpressurized fluid 
\cite{Lach,Byefd,Riceaa,bnflal,slelokcfla,mfma}.
It is more and more recognized that fluids play an essential role in virtually all
crustal processes. Ref.\cite{hivk}
reviews the historical development
of the conciousness among researchers of the ubiquitous presence and importance
of fluids within the crust.
Numerous examples exist that demonstrate water as an active agent
of the mechanical, chemical \cite{wyufj}
and thermal processes that
control many geologic processes that operate within the crust \cite{com,myurev}.
The bulk of available information on the behavior of fluids comes
from observations of exposed rocks that once resided at deeper crustal levels. 
In any case, present day surface exposed metamorphic rocks indicate that, at all crustal
levels, fluids have been present in significant volume. Because the
porosity of metamorphic rock is probably less than $1 \%$, the high volume of calculated
fluid necessary to produce observed chemical changes suggests that fluid must have
been replenished thousands of times. 
It has also been proposed that gold-quartz vein
fields in metamorphic terranes provide evidence for the involvement of large volumes
of fluids during faulting and may be the product of seismic processes 
\cite{rijmds}. Water is also released from transformed minerals. 
For instance, Montmorillonite changes to illite
with a release of free water from the clay structure at approximately the same depth
as the first occurrence of the anomalous pore pressure 
\cite{mhshs}. This is
the most commonly discussed example of hydration and dehydration of minerals changing
the fluid mass and the pore pressure.
Fluids have been directly sampled
at about $11 ~km$ by the Soviets at the Kola Peninsula drillhole. 

Many observations suggest that there are massive crustal fluid
displacements correlated with seismic events. Among them, one can cite the
fault-valve mechanism \cite{mskjsa} or the migration and diffusion of
aftershocks \cite{nurbok}. 
Several mechanisms have been
proposed for the mechanical effect of fluids
to decrease compressive lithostatic stresses:
mantle-derived source of fluids can maintain
overpressure within a leaky fault \cite{Riceaa};
laboratory sliding experiments on granite show that the
sliding resistance of shear planes can be significantly decreased by pore sealing
and compaction which prevent the communication of fluids between the porous
deforming shear zone and the surrounding material \cite{bnflal}.
Physico-chemical processes such as
mineral dehydration
during metamorphism may provide large fluid abundance
over large areas \cite{shchos,myurev}.
Such fluid presence or migration implies that cracks
may open and close even at seismogenic depth, justifying the relevance of
asymetric nonlinear elasticity for such depths as we elaborate below.

\subsection{Physical mechanism for the threshold nonlinearity}

From a mechanical point of view, the threshold nonlinearity we invoke
is different from the ubiquitous nonlinearity
of rocks. In the later, nonlinearity becomes important only under large deformations.
Such nonlinear elasticity of rocks is
well-documented from its nonlinear wave signatures \cite{mskjks}.
In contrast, the nonlinearity we invoke is revealed for almost arbitrary small
perturbations by the difference between compressive versus extensive
perturbations around a mean lithostatic stress field. 
Many crustal rocks have
a Young's modulus depending on confining pressure, in particular with
a Young's modulus in tension smaller than the Young's
modulus in compression in a ratio from $1/2$ to $1/10$ \cite{jsjs}.

The highly damaged upper crust, as described above, would behave as a standard continuous elastic 
medium if almost
all cracks are kept closed and are thus completely transparent to the applied tectonic
stress. In a totally dry crust, this would occur at depths larger than about $1km$.
Near the surface, cracks can open under sufficiently tensile stresses, so that the rock 
(if we
neglect crack growth phenomena) will display low elastic modulii (what we will call
later a {\it soft} state). Under compressive (or weakly tensile) stresses, those same 
cracks will close
so that the rock will display elastic modulii close to those of uncracked material
(the {\it hard} state). This simple compressive/tensile asymmetry changes strongly
the behaviour of the rock: the rheology is still elastic (in the sense that it is
reversible) but it is nonlinear (a change of sign of applied stress does not only changes
the sign of strain, it also changes its modulus). In addition, if we now take account
of the possible slow growth of cracks under tensile stress, the nature of the 
nonlinearity will not change, but its amplitude will.

Let us now consider the influence of fluids whose presence are pervasive in the crust,
as summarized in section \ref{jfgkal}.
Let us assume that the crust is saturated with fluids and that the network of cracks
is sufficiently dense so that it behaves as a drained medium. For the sake of 
simplicity, we will also assume that the crust is characterized by a homogeneous
spatial distribution of cracks, so that the permeability is uniform and
isotropic. We will also suppose that there exists a horizontal interface
at depth $z_{seal}$ where permeability is $0$. The last ingredient is 
that the part of the crust below $z_{\rm seal}$ is connected to a reservoir of fluid.
Many indirect observations suggest indeed the presence of large sources of fluids
\cite{Wannamkaer,Kirby}.
Then, above depth $z_{\rm seal}$, water will be at hydrostatic pressure $P_h$ (lower than
lithostatic
pressure $P_l$) so that cracks will be able to close if the tensile applied tectonic 
stress is smaller than $P_l - P_h$ in modulus (cracks remain closed if tectonic stress is 
compressive). Below $z_{\rm seal}$, the scenario is different. Water trapped in cracks is now
at lithostatic pressure. If there is no applied tectonic stress, fluid pressure inside 
cracks compensates exactly the lithostatic pressure and the medium is exactly at the 
hard/soft boundary. The net stress acting on any crack's lips is $0$, so that the crack
doesn't grow.
If the applied tectonic stress is compressive, cracks will close and fluids are expelled
towards the reservoir: the crust is in the hard phase. If the applied tectonic stress 
is tensile, cracks will open and the crust is in the soft phase. If a crack becomes
unstable, pressure drops within it, so that it tends to close to re-establish initial
fluid pressure. If a crack grows very slowly, pressure within the crack will stay constant
with time. We will from now on neglect the possible slow crack growths and assume that fluid 
pressure
is constant in each crack and that the crack network geometry does not show any 
evolution with time or applied stress. Thus, below
$z_{\rm seal}$, the applied stress threshold to get from the hard to the soft state is $0$. 
This explains our terminology of a ``threshold nonlinearity.''

The remaining of the paper will investigate stress
redistributions associated with earthquakes occurring below $z_{\rm seal}$. The existence
of $z_{\rm seal}$ is difficult to prove for the real crust. It would correspond, for
instance, to a depth where the crack network geometry changes abruptly, for instance
at the boundary between the sedimentary basin and cristalline rocks. We can also
propose an alternate model in which the vertical permeability below $z_{\rm seal}$ is
much lower than the horizontal one, and would obtain the same kind of soft/hard 
transition. Note that much more complicated scenarii involving the chemistry of both
fluids and rock matrix could be taken into account \cite{myurev}, but we choosed to neglect them
in this purely mechanical paper.

We then propose a simple central-force spring model in two dimensions
to study the behaviour of such an asymmetric nonlinear medium when
subjected to an infinitesimal internal source of stress or strain.
We show that the spatial structure of the stress transfer associated
with such a perturbation evolves with the strength of the nonlinearity
(defined as the ratio of the springs' stiffness in tensile and in compressive
states). Those changes are quantified in terms of the symmetry of the resulting
stress field and of the decay rate of the amplitude of the stress 
perturbation with distance from the source.

\section{Numerical Model}

Solving theoretical problems of nonlinear elasticity proves to be very tough, even
for the simple asymmetry of our threshold nonlinearity. We thus choosed to solve a couple of
simple problems related to seismology using numerical modelling. Stress, strain and
material rigidity being $2^{nd}$ or $4{th}$ rank tensors, and as there is an obvious and 
complex feedback
between strain and rigidity in our model, we choosed to use a simple spring model
to illustrate the concept and major consequences of the threshold nonlinear elasticity.

\begin{figure}
\epsfig{file=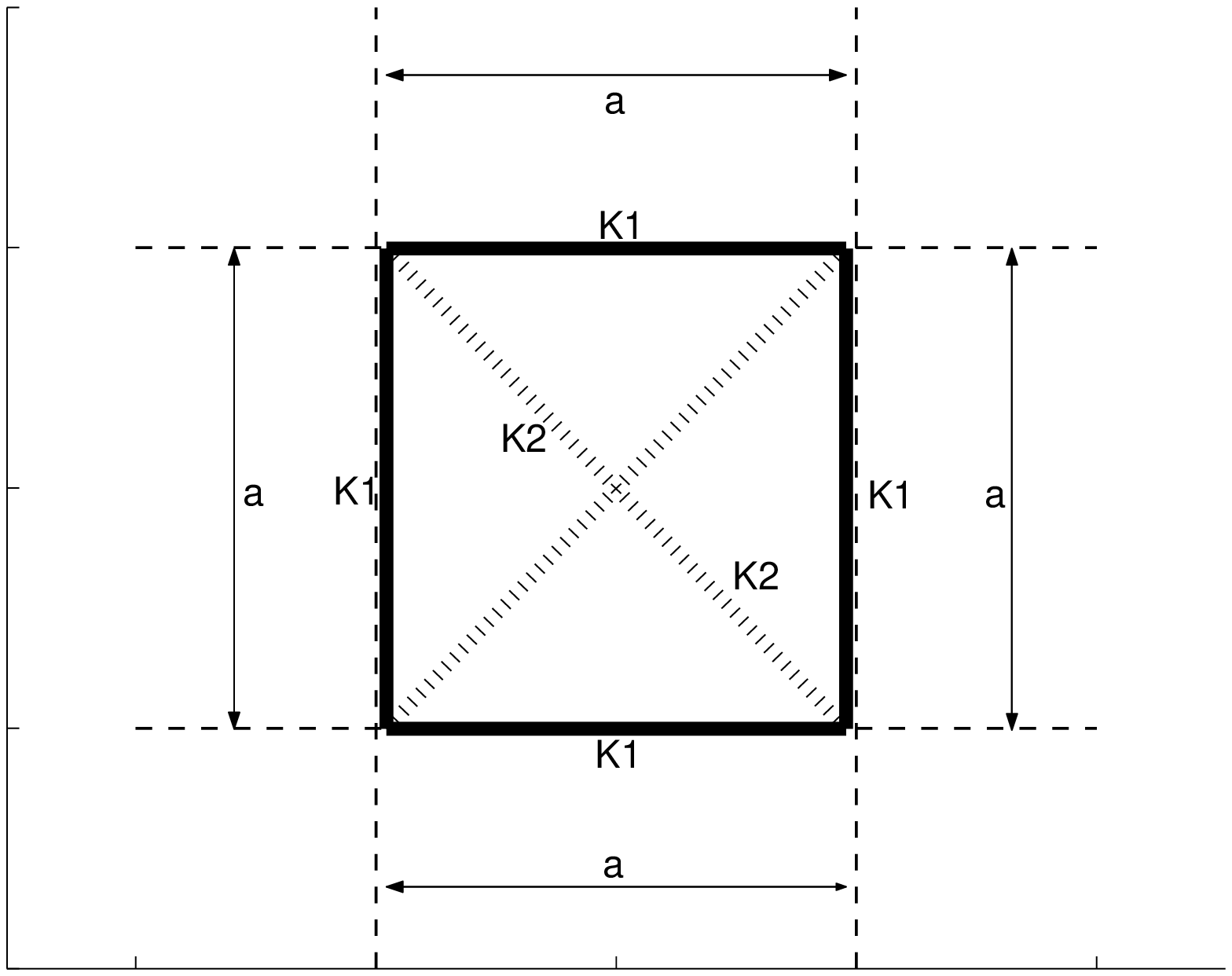,width=16cm}
\caption{\label{defcell} Mechanical structure of an elementary cell of the model.
Each cell is defined by $4$ nodes linked by springs along edges and both diagonals.
Springs on the edges are of stiffness $K_1$, while diagonal springs are of stiffness
$K_2=K_1/2$ (see text).}
\end{figure}

A plate of size $L$ by $L$ is discretized onto a regular grid of mesh size $a$. 
In the following, we choose $L=2000km$ and $a=10km$.
Each elementary cell is defined by $4$ nodes, each node being shared between $4$ 
different neighbouring cells (except on the boundary of the plate). 
Figure \ref{defcell} shows the mechanical structure defined for each cell: each
node is connected to its two nearest neighbours by springs of stiffness $K_1$ (those
springs indeed define the $4$ edges of the cell). As those are central force springs,
the shear modulus of such a cell is indeed $0$. To get shear elasticity \cite{Attila},
$2$ springs are added along
the cell's diagonals, such that each node is also connected to its next-nearest neighbours.
The stiffness of those diagonal springs is $K_2$. Once the plate is discretized with such
cells, it can be shown that the plate behaves as an isotropic elastic medium if and only 
if
we have $K_2 = K_{1}/2$ \cite{Monette}. The two independent elastic modulii of the plate can 
then be shown to be $\lambda=\mu=K_1/2$
for the two Lam\'e coefficients, thus yielding $E=(4/3) K_1 = (8/3) \mu$ for the Young modulus 
and $\nu=1/3$ for the Poisson coefficient (note that we are dealing with a pure
$2D$ model). Of course, many other geometries of the elementary cell are possible, but
we had to choose square cells which help to handle more easily with boundary conditions 
used in the problems we want to study.

The relationships we just defined are true only if the springs are symmetric, i.e.,
their stiffnesses is the same under tensile or compressive states. The next and last
step to model our nonlinear threshold elastic rheology is to impose that the stiffness of
each spring can vary with its length. Thus, if a spring is shortened, its stiffness
will be, say, $K$. If a spring is lengthened, its stiffness is lower and taken equal to
$\alpha K$, with $\alpha \leq 1$. Each spring represents an effective volume filled
with a uniform and 
isotropic distribution of cracks with sizes smaller than the
representative mesh size. Real damage in the crust is of course
much more complex, with anisotropic, space and scale
dependence. These complications are neglected in our first
exploration. We can obtain an order of magnitude
estimate of the density of fluid-filled faults associated with
a given asymmetric coefficient $\alpha$, using
the effective medium calculations in
\cite{GarbinKnopoff}. To simplify, let us assume
$\lambda=\mu$. Then, $\alpha = \lambda_d/\lambda_1 = 1/(1+ 5d/2)$,
where $\lambda_d$ is the Lam\'e coefficient of the damaged material
and $d = N (\ell/L)^3$ is the density of faults (assumed identical)
of radius $\ell$ in a  cube of volume $L^3$. $N$ is the number of faults
in that volume. For instance, we need about $11$ faults of size $L/3$
to get $\alpha=0.5$. Such estimate must however be taken with caution
since the effective medium calculation $\alpha =1/(1+ 5d/2)$
is valid only for small crack densities, while any piece of rock
and the real crust are crisscrossed by many faults at many length scales,
most of them being healed at varying degrees. We think that values
of $\alpha$ significantly smaller than $1$ should thus not be excluded.
It is also probably that $\alpha$ is not uniform within the crust
and can be expected to reflect the past history of deformations and ruptures.

The ratio $\alpha$ of the extensive over compressive
elastic coefficient can also be seen as equivalent to $1-D_m$ in damage
mechanics, where $D_m$ is the scalar damage variable. If the spring isn't damaged, then 
$D=0$, so that $\alpha=1$ and the stiffness is the same under tensile and compressive
states. If the spring is totally damaged (near failure), $D_m$ is close to $1$, so that
$\alpha \simeq 0$ and the stiffness of the spring in tensile state vanishes, while
it is still $K$ if the spring is compressed. Under arbitrary loading conditions, some
springs in the plate will be in tensile state, while others will be in compressive state,
so that it is difficult to analytically compute the stiffness tensor of the whole plate
when $\alpha < 1$. In addition, the stiffness tensor may feature more than $2$ 
independent modulii. This justifies the use of an iterative numerical method as described
in the next section.

\section{Numerical Method}

The method we use belongs to the so-called iterative `type-writer' methods.
The first step consists in defining boundary conditions, i.e. to fix displacements
and/or applied forces on set of nodes. 
As we are dealing with statics, such forces and displacements are held constants 
throughout the numerical process.
We then consider, say, the node on the top left corner of the mesh and, according
to the forces/displacements applied to this node and his nearest and next-nearest
neighbours, we can compute the net force acting on that node. As we are dealing with
a statics problem, the net force acting on that node at equilibrium should vanish.
If the force vector we computed for that node is not $\overrightarrow{0}$, then we
move it in the direction of the force vector to decrease the net force. We store the
new position of this node and get to its right neighbour and follow the same scheme.
We thus sweep the mesh line by line down to the node on the bottom right, and iterate
the same operations again from the node on top left. As iterations accumulate, the net
force acting on each node decreases, and we stop the process once all nodes are
subjected to a force whose modulus is under a given threshold. This threshold
is chosen such that the modulus of the incremental displacement necessary to decrease
the net force modulus on each node is of the order of the accuracy of the computer,
namely about $10^{-12}$. To avoid any spurious result due to the direction of type-writing,
iterations alternatively begin on each of the four corners of the plate on either 
horizontal or vertical directions, right or left, up or down.
Once the `type-writer' is stopped, displacements of all nodes relative to their initial
positions are stored. Those positions also allow to compute, for a given node, the forces
exerted on it by each of its neighbours. Those last quantities allow to compute the full
stress tensor at that node \cite{Monette}, which is also stored.
Computations of forces transmitted by a spring from one node to another take account
of the spring stiffness which, as described in the previous section, depends on the
state (stretched or shortened) of the spring through the value of $\alpha$ which is
kept constant throughout the network.

Starting from given boundary conditions, we solve a statics problem, which means that 
we do not take account of neither wave propagation nor fluid migration. We just compute the
final equilibrium solution. This point will be discussed at the end of the paper when
considering the application of this model to real Earth data.

\section{Earthquake Modeling}

In this paper, we want to examine the main differences between the stress
field pattern generated by an earthquake in our nonlinear threshold elastic medium and in a 
standard linear elastic medium. The first step is thus to define mesh parameters
representative of the real crust, the second one is to define what is an earthquake
source in such a model.

\subsection{Parametering the Earth's crust}

Our numerical model is strictly a $2D$ one, as dealing with the third dimension
would lead to serious memory and computing time problems, and we would have to handle
boundary problems such as the free surface and coupling with lower viscous layers.
We can however choose
mesh parameters such that, using relationships linking plane elasticity to $3D$ 
elasticity, we can model realistic crust properties (however neglecting boundary problems).
The size of the plate as well of the cells has been previously given. We fixed the 
(virtual) thickness of the plate to be $10km$, so that it corresponds roughly to the 
thickness of
the seismogenic zone within the crust. We then choosed $K_c=5 \cdot 10^{14}~Nm^{-1}$ 
(where $K_c$ is the value of $K_1$ when springs are compressed), so that the $3D$ elastic
modulii become $E=6.25 \cdot 10^{10}~Pa$, $\lambda=\mu=2.5 \cdot 10^{10}~Pa$, and $\nu=0.25$,
which are close to usual modulii measured in rock mechanics experiments. 
The value of the assymmetry parameter $\alpha$ will be varied through several numerical 
experiments from $1$ (standard isotropic elasticity) down to $0.01$ (strong asymmetry
of the elastic response under extension versus compression).

\subsection{Earthquake Source modelling}

Earthquake source theory has until now been theoretically studied within
the framework of linear elasticity. This allows one to use very powerful
tools such as the representation theorem and Green functions. An earthquake
can then be viewed equivalently as a displacement discontinuity across a
fault plane, or a distribution of double-couples and dipoles of forces along
the same plane in a continuous medium \cite{Aki,Burkno,Pujol}.

In the case of a fault of finite dimensions, if we assume that the stress drop
is uniform along the fault, then we deal with a crack problem. If we assume that
the displacement discontinuity across the fault is uniform, then we deal with
a dislocation problem. At distances from the fault much larger than its size, 
and in the case of linear elasticity, both
models yield the same spatial patterns of stress and displacement fields, which are
linked by Hooke's law. This will be illustrated below in our diagrams obtained
for the symmetric elasticity case $\alpha=0$.

In the nonlinear case, it is easy to show that the representation theorem 
fails to apply,
and then so does the Green function concept. This stems from the fact that 
the principle of linear superposition fails in the presence of nonlinearity.
It follows that even the most
simple earthquake source problem has to be defined either as a crack
or a dislocation problem, and both problems should
give different stress and displacement patterns at long wavelengths. 
We will, in this preliminary
work, study only pointwise sources, which will allow simple comparisons with elementary
solutions obtained from linear elasticity.

We will thus study two cases: 
\begin{itemize}
\item[(i)] an initially stress-free medium within which
a single cell (located at its center) is subjected to the following stress field tensor: 
$\sigma_{xx}=\sigma_{yy}=0$
while $\sigma_{xy}\not=0$ (and is hereafter called {\it pointwise shear stress load} or
crack model) - this model is reminiscent of the standard dynamical model of an event in standard
linear elasticity \cite{Aki},
\item[(ii)] an initially stress-free medium within which a
single cell is subjected to a pure shear strain field: $\epsilon_{xx}=\epsilon_{yy}
=0$ while $\epsilon_{xy}\not=0$ (hereafter called {\it pointwise shear strain load} or 
dislocation problem) - this model rather views the event as a shear displacement discontinuity.
\end{itemize}
In both cases, the corresponding infinitesimal planar defect (the source of the
earthquake) suffers from undeterminacy
and is oriented either along the $\vec{O}_x$ direction (plane $P_x$) or along
the $\vec{O}_y$ direction (plane $P_y$). In the first case, the slip discontinuity is dextral,
and it is sinistral in the second case.

\subsection{Quantitative source parameters}

\begin{figure}
\epsfig{file=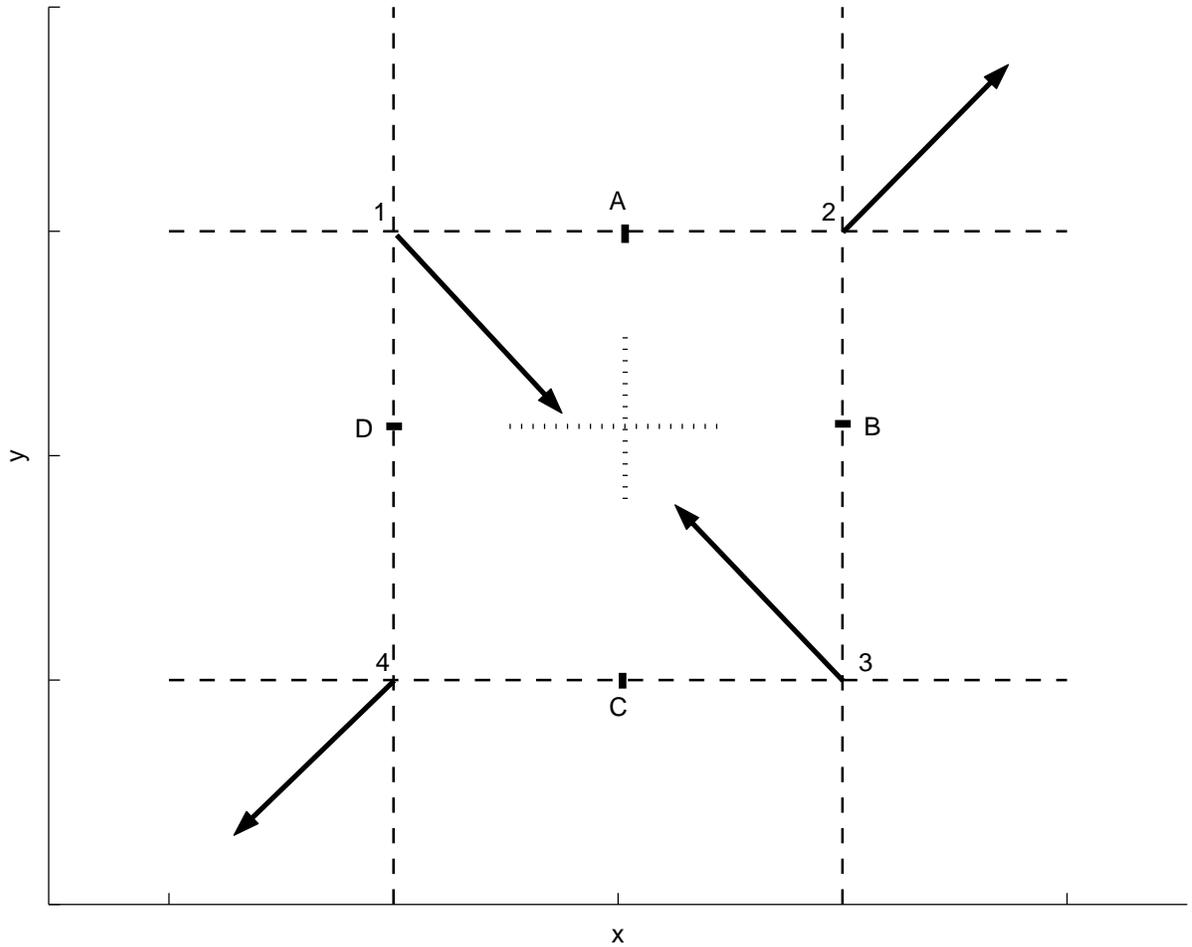,width=16cm}
\caption{\label{forcedispsource} Representation of the elementary source in 
a unit cell.
Vectors represent either forces of equal modulii (crack model) or displacement of
same amplitude (dislocation model). Nodes of the cell are labelled $1,2,3,4$. 
Dotted lines represent the virtual plane defect along which force or displacement
discontinuities are imposed to model an earthquake. Those discontinuities have to 
be computed at points labelled $A,B,C,D$ (see text).}
\end{figure}

The small scale of our mechanical model is that of a cell, and this is thus
the smallest scale we have to deal with in order to model an earthquake source.
Figure \ref{forcedispsource} shows the source cell and $4$ different vectors
originating from each of its $4$ corners.

In the case of the pure shear stress load model, each vector represents a force applied
to the corresponding node. All forces have the same modulus so that the stress tensor
within the central cell indeed corresponds to the one we defined above. The same set of
forces is applied whatever the value of $\alpha$.

In the case of the pure shear strain load model, each vector represents a displacement
applied to the corresponding node. All displacements have the same modulus so that the
strain tensor within the central cell indeed corresponds to the one we defined above.
The same set of displacements is applied whatever the value of $\alpha$.

In order to ensure that we can compare results obtained from both types of boundary 
conditions, we have to fullfil a very simple condition: the stress and displacement
fields must be identical in the classic linear case, i.e. when $\alpha=1$. 
In the pure shear stress load model, we imposed the modulus of each applied force
equal to $F=7.1 \cdot 10^{14}N$, so that is corresponds to an event of scalar moment 
$M_0 = 7.1 \cdot 10^{18}~Nm$, i.e., of magnitude $\approx 6.5$. We computed the 
displacement field for $\alpha=1$, and observed that the magnitude of the displacement
at each node of the source cell was $0.526640588 m$. To be perfectly consistent, we thus 
impose this displacement amplitude at each source cell node in the case of
the pure shear strain load model.

\section{Displacement field at the source}

In the pointwise dislocation case, displacements are held constant at the source 
whatever the value 
of $\alpha$. In the pointwise crack model, only forces are kept constant as $\alpha$ is
changed,
and displacements are expected to vary as the asymmetry increases (i.e. $\alpha$ decreases).
We have already pointed out that both boundary conditions 
assume that the mechanical defect is either
parallel to $P_x$ or to $P_y$. Indeed, we will show that the displacement 
(hereafter named $u_n$) normal to each
of these conjugate defects is of the same type (opening) while 
the shear displacement (hereafter $u_s$)
along each of them is of different type: dextral along $P_x$ and sinistral along $P_y$.
That said, we will focus only on the modulii of those displacements.

How do we obtain $u_n$ and $u_s$? Indeed, according to the orientations of the conjugate
plane defects, we have to compute displacements at points $A,B,C$ or $D$, whereas our
model provides solutions at nodes $1,2,3$ and $4$ (see Figure \ref{forcedispsource}). Displacements
at points $A$ to $D$ are thus computed through bilinear interpolations within the cell.
We then define $u_n=u_x(B)=-u_x(D)=-u_y(C)=u_y(A)$ and $u_s=u_x(A)=u_y(B)=-u_x(C)=-u_y(D)$
as a result of the symmetries of the system.

\begin{figure}
\epsfig{file=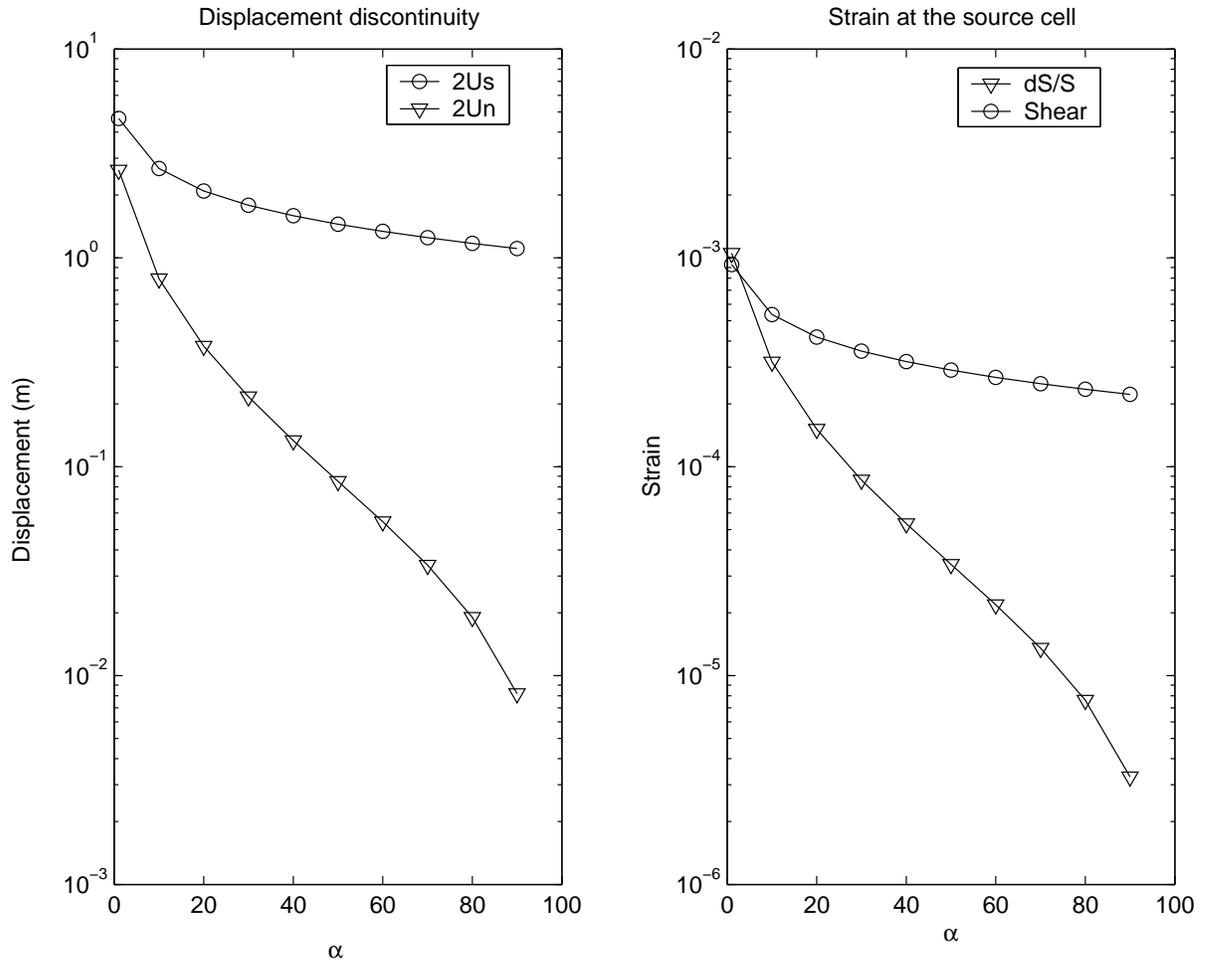,width=16cm}
\caption{\label{u_n_u_s_alpha} Left panel: shear (circles) and normal (triangles) 
displacements discontinuities along planar defects within the source cell as a 
function of $\alpha$
in semilogscale. Results for $\alpha=0$ are not shown as the normal discontinuity
vanishes. Right panel: volumetric (circles) and shear (triangles) strains of the source
cell as a function of $\alpha$ in semilogscale ($\alpha$ is the ratio of the 
elastic modulus in extension over the elastic modulus in compression). 
Results for $\alpha=0$ are not shown as 
the volumetric strain vanishes in this case. 
All results are obtained for the pure shear stress loading case.}
\end{figure}

Figure \ref{u_n_u_s_alpha} shows the variations of both $u_n$ and $u_s$ with $\alpha$
(in fact, it shows the values of displacements discontinuities across the crack, i.e. $2u_n$
and $2u_s$).
For $\alpha=1$, we find that $u_n$ is very close to $0$, so that the displacement along
the pointwise crack lips is pure shear. As $\alpha$ decreases from $1$ to close to
$0$, the
shear displacements increase by a factor of about $4.5$. This is perfectly understandable,
as some springs are sollicited in tension, leading to a decrease in their stiffness. This 
decrease, in the presence of constant forces, implies that displacements increase.

More surprising is the behaviour of normal displacements, which increase drastically as
$\alpha$ decreases, tending to be about $2.7m$ when $\alpha$ tends to $0$ (i.e. about half
the shear displacements). For $\alpha=0.1$ we have $u_n/u_s$ close to $1/3$. 
Moreover, $u_n$ and $u_s$ are both positive, which signifies
that, under the shear stress load assumption, the defect opens when the medium is 
asymmetric. In seismological words,
this means that the static moment tensor of the source has a non-vanishing trace and
can thus be decomposed into an isotropic part and a deviatoric one. Despite the
observation that most
earthquake sources are thought to be well modelled by the deviatoric part alone, a few 
catalogues report isotropic components. Several mechanisms have been invoked to explain
a non-vanishing isotropic component of the seismic moment.
The standard explanation for 
non-double-couple components relies on the fault zone irregularity \cite{Kuge}.
Some earthquakes with non-double-couple mechanisms have been claimed
not to be explained solely by such a
composite rupture \cite{ndcfro,myurev}.  
It is then important to note
that the seismic moment tensor reported in catalogues is a variable that quantifies
{\it static displacements} at the source. A static dilational strain
at the source can occur even when the dynamic representation of the source is a pure
double couple of forces (yielding a stress tensor with zero trace).

Figure \ref{u_n_u_s_alpha} also reports the shear strain that can be measured at the source 
cell as well as the relative dilation of its surface $dS/S$. Both quantities of course
corroborate the previous results on $u_s$ and $u_n$.

\section{The stress field}

We will now focus on the structure of the stress field generated by our pointwise
earthquakes. The stress field will be studied at large wavelengths, i.e. at distances
from the source cell of more than a few cell sizes. This gives
the rate of stress decay with distance from the source and thus the range of
interactions between events. We will in
the following implicitly assume that the plate is affected by many other faults that are
locked and oriented in the same direction (say, along direction $P_x$ with potential dextral
displacement along the fault). The source
cell is one of such fault producing an event. We then study the effect of this event
on all other faults in the plate.

\subsection{Spatial patterns}

\begin{figure}
\epsfig{file=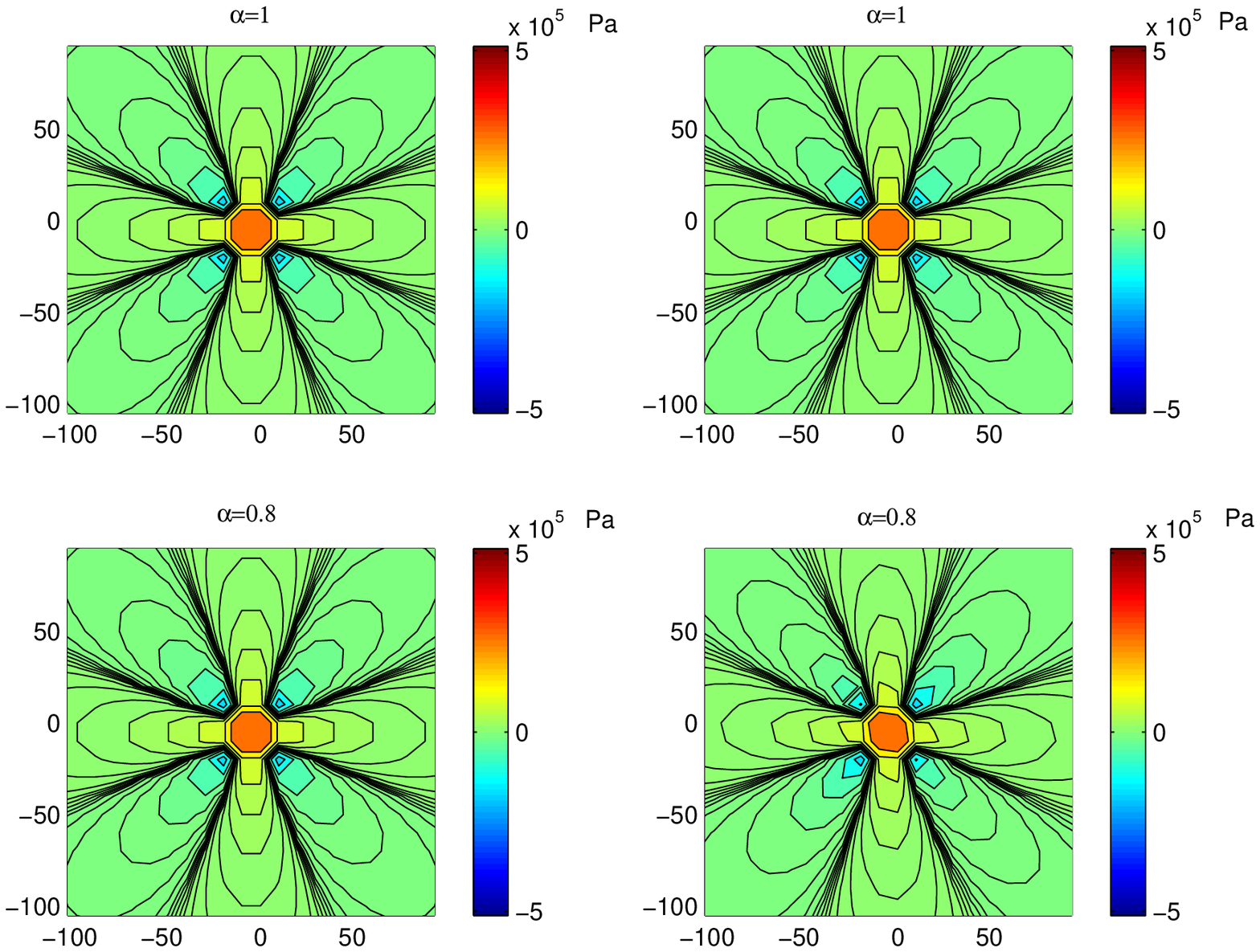, width=16cm}
\caption{\label{sxy_100_80} Map of the shear stress transfer
$\sigma_{xy}$ for $\alpha=1$ and
$\alpha=0.8$ in the pure shear stress load case (left panels) and in the pure shear
strain loading case (right panels). The source is located at $(0,0)$ and all space
units are kilometers.}
\end{figure}

\begin{figure}
\epsfig{file=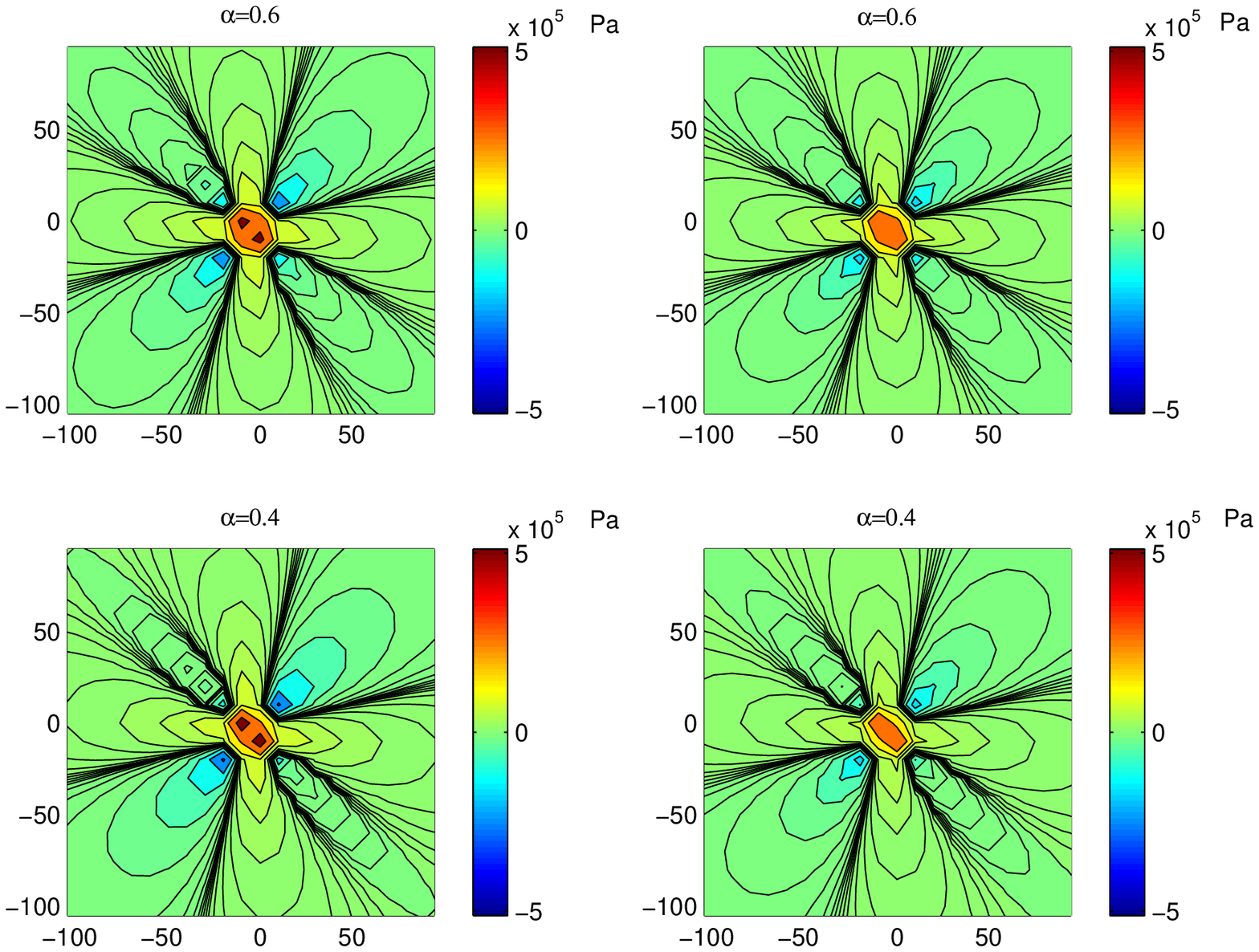, width=16cm}
\caption{\label{sxy_60_40} Same as Fig.~\ref{sxy_100_80}
for $\alpha=0.6$ and $\alpha=0.4$.}
\end{figure}

\begin{figure}
\epsfig{file=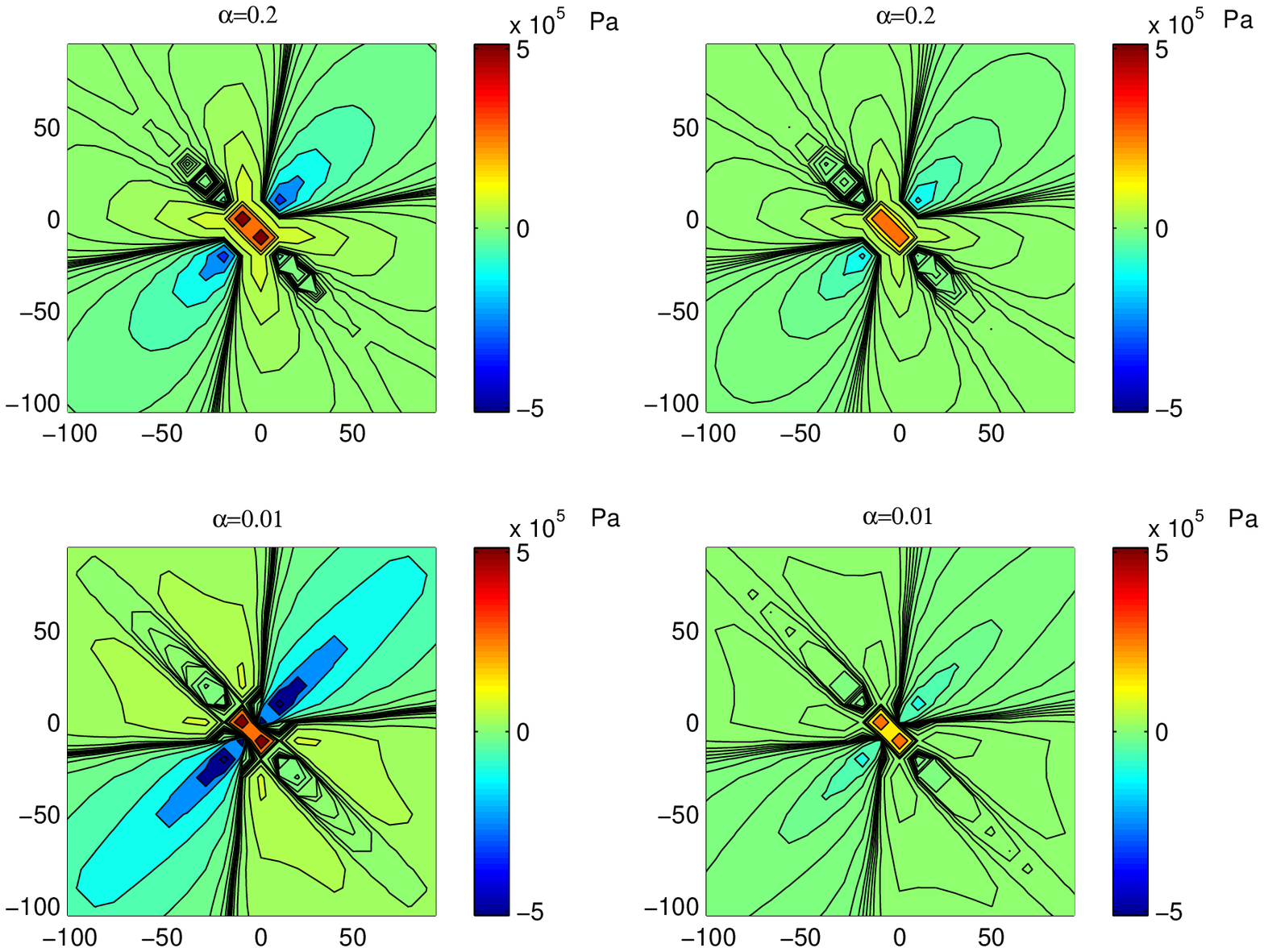, width=16cm}
\caption{\label{sxy_20_1} Same as Fig.~\ref{sxy_100_80}
for $\alpha=0.2$ and $\alpha=0.01$.}
\end{figure}

Figures \ref{sxy_100_80} to \ref{sxy_20_1} show the variation
of the shear stress component $\sigma_{xy}$ within the plate near the source cell. 
A positive variation signifies that this
stress component increases. On each figure, the panels on the left represent patterns
obtained with the pure shear stress load hypothesis, while the panels on the right
represent patterns obtained with the pure shear strain load hypothesis. Each row
corresponds to a different value of $\alpha$, i.e., to a different degree of 
elastic asymmetry between extensive and compressive deformation.

In the case $\alpha=1$, both boundary conditions yield exactly the same spatial 
pattern, as expected. This is in agreement with the fact that, in linear elasticity, both boundary
conditions are equivalent. We obtain $8$ lobes of identical shapes within which stress
amplitude alternates from positive to negative (red and blue lobes, repectively).

As $\alpha$ decreases, the symmetry of the patterns decreases: some lobes are rotating,
some are widening while others are narrowing. We will study stress variation in those
lobes in a subsequent section. The most striking observation is however that the spatial
structure of the patterns is the same for both types of boundary conditions for a given
value of $\alpha$, notwithstanding a significantly smaller stress amplitude in the pure
shear strain load boundary condition. 

\begin{figure}
\epsfig{file=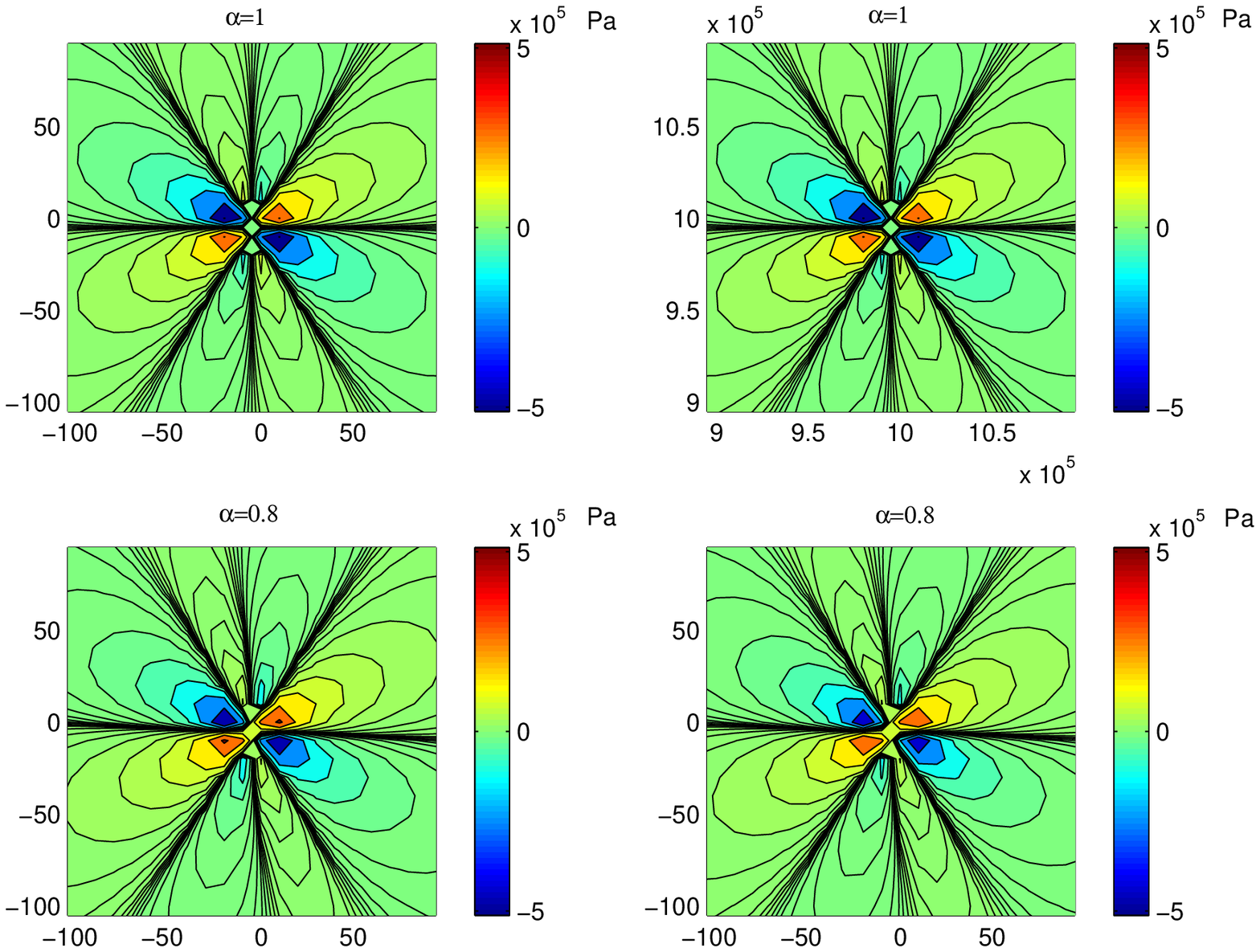, width=16cm}
\caption{\label{sxx_100_80} Map of the stress transfer $\sigma_{xx}$ for $\alpha=1$ and
$\alpha=0.8$ in the pure shear stress load case (left panels) and in the pure shear
strain loading case (right panels). The source is located in $(0,0)$ and all space
units are kilometers.}
\end{figure}

\begin{figure}
\epsfig{file=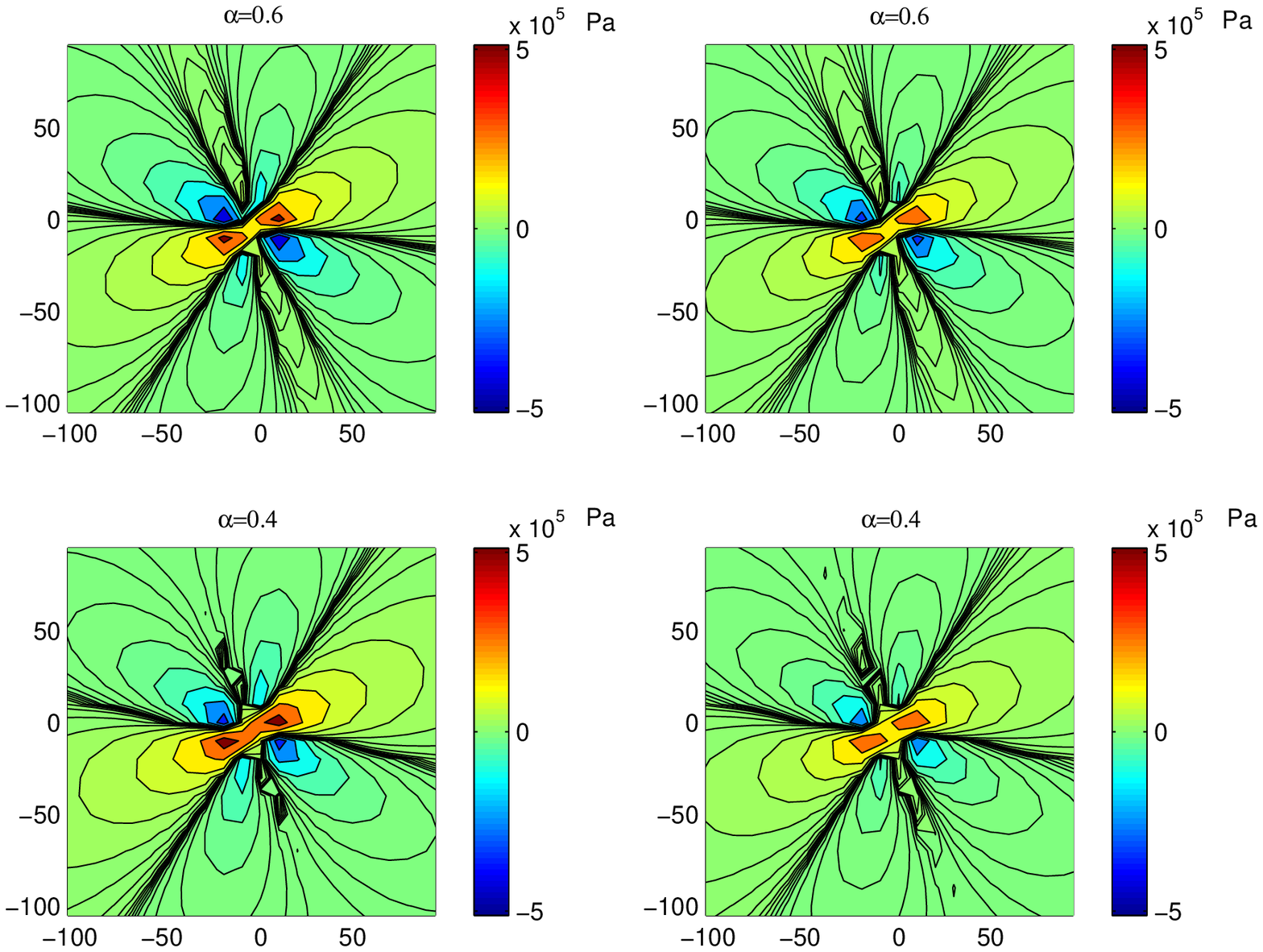, width=16cm}
\caption{\label{sxx_60_40} Same as Fig.~\ref{sxx_100_80}
for $\alpha=0.6$ and $\alpha=0.4$.}
\end{figure}

\begin{figure}
\epsfig{file=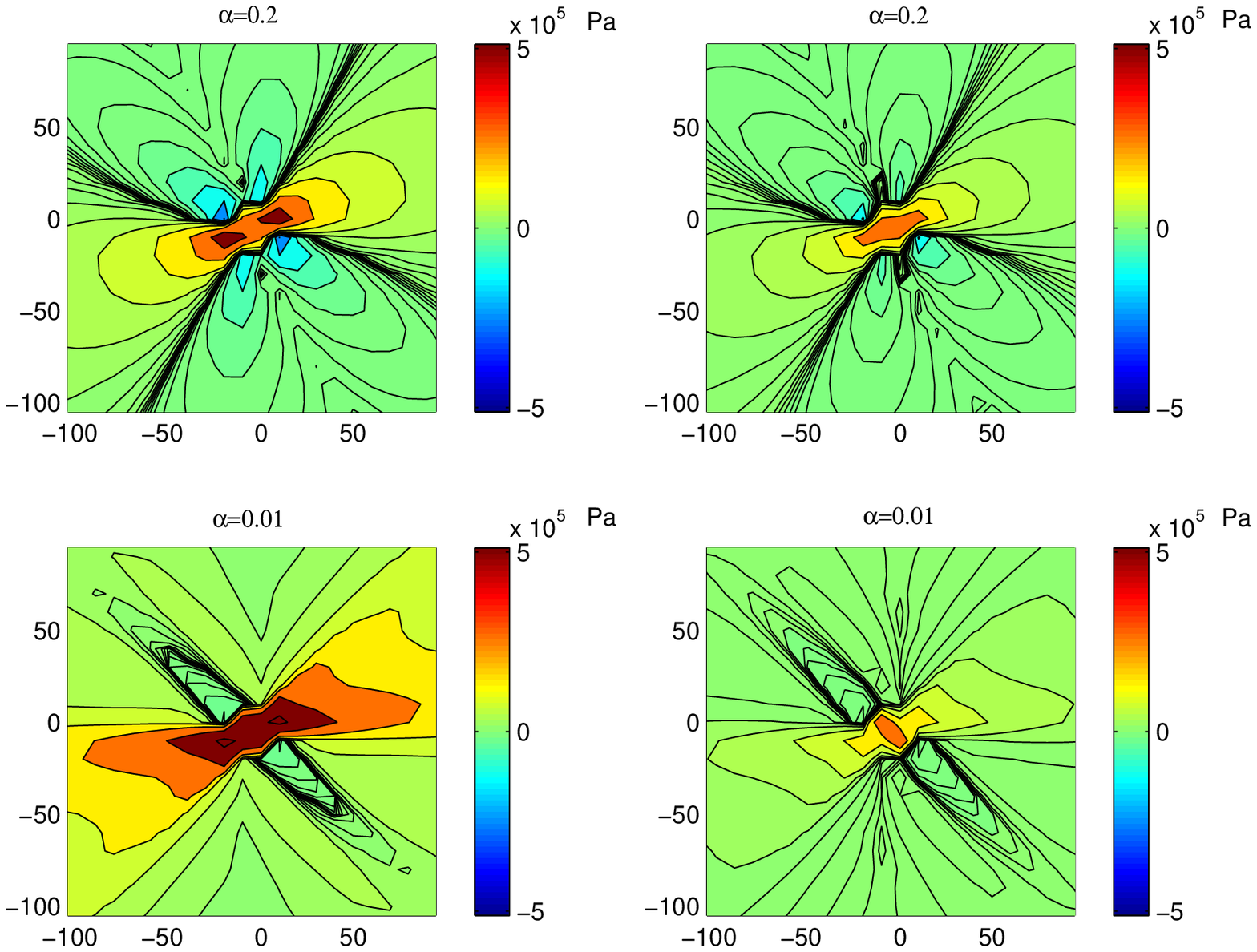, width=16cm}
\caption{\label{sxx_20_1} Same as Fig.~\ref{sxx_100_80}
for $\alpha=0.2$ and $\alpha=0.01$.}
\end{figure}

\begin{figure}
\epsfig{file=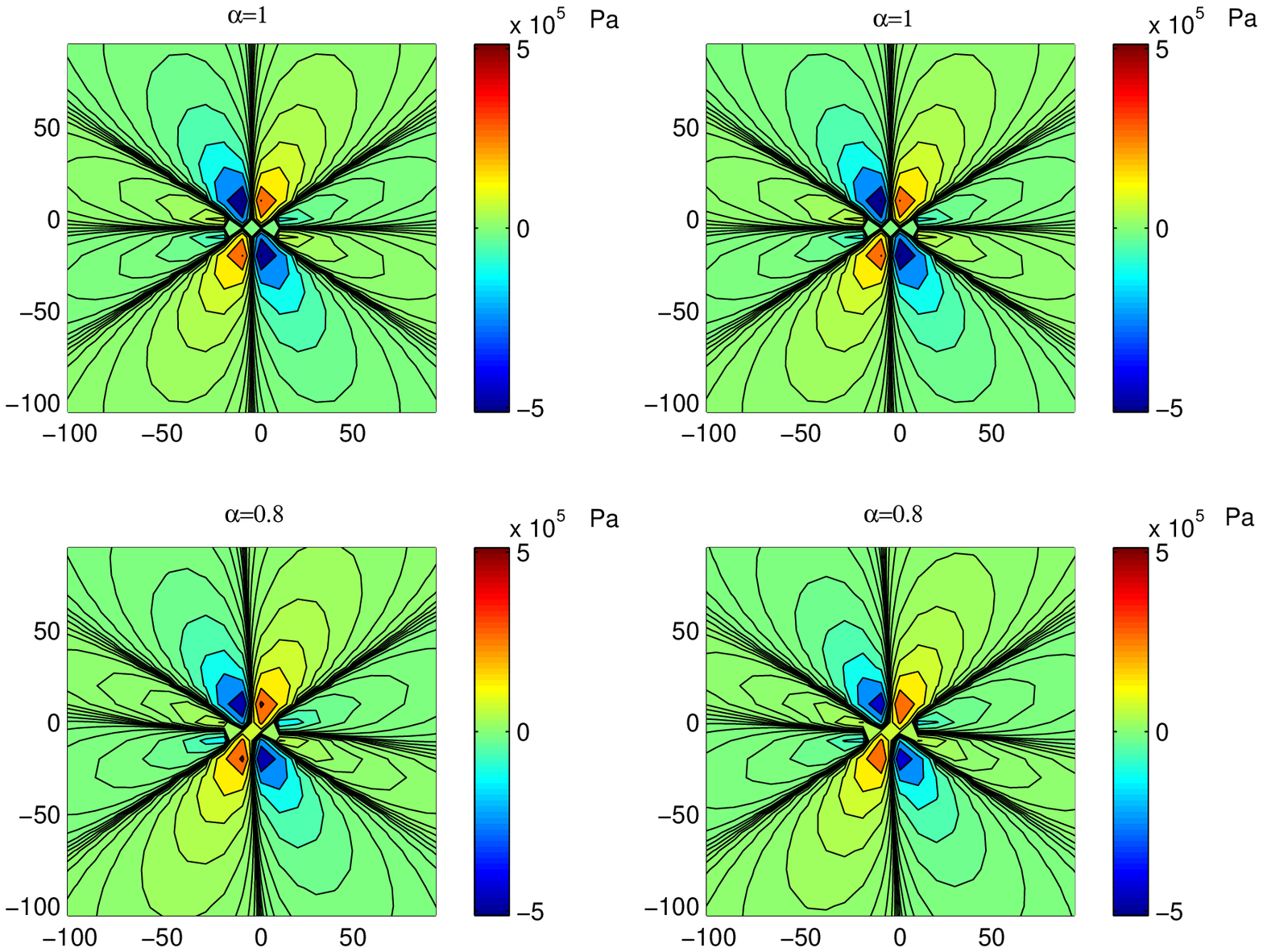, width=16cm}
\caption{\label{syy_100_80} Map of the stress transfer $\sigma_{yy}$ for $\alpha=1$ and
$\alpha=0.8$ in the pure shear stress load case (left panels) and in the pure shear
strain loading case (right panels). The source is located in $(0,0)$ and all space
units are kilometers.}
\end{figure}

\begin{figure}
\epsfig{file=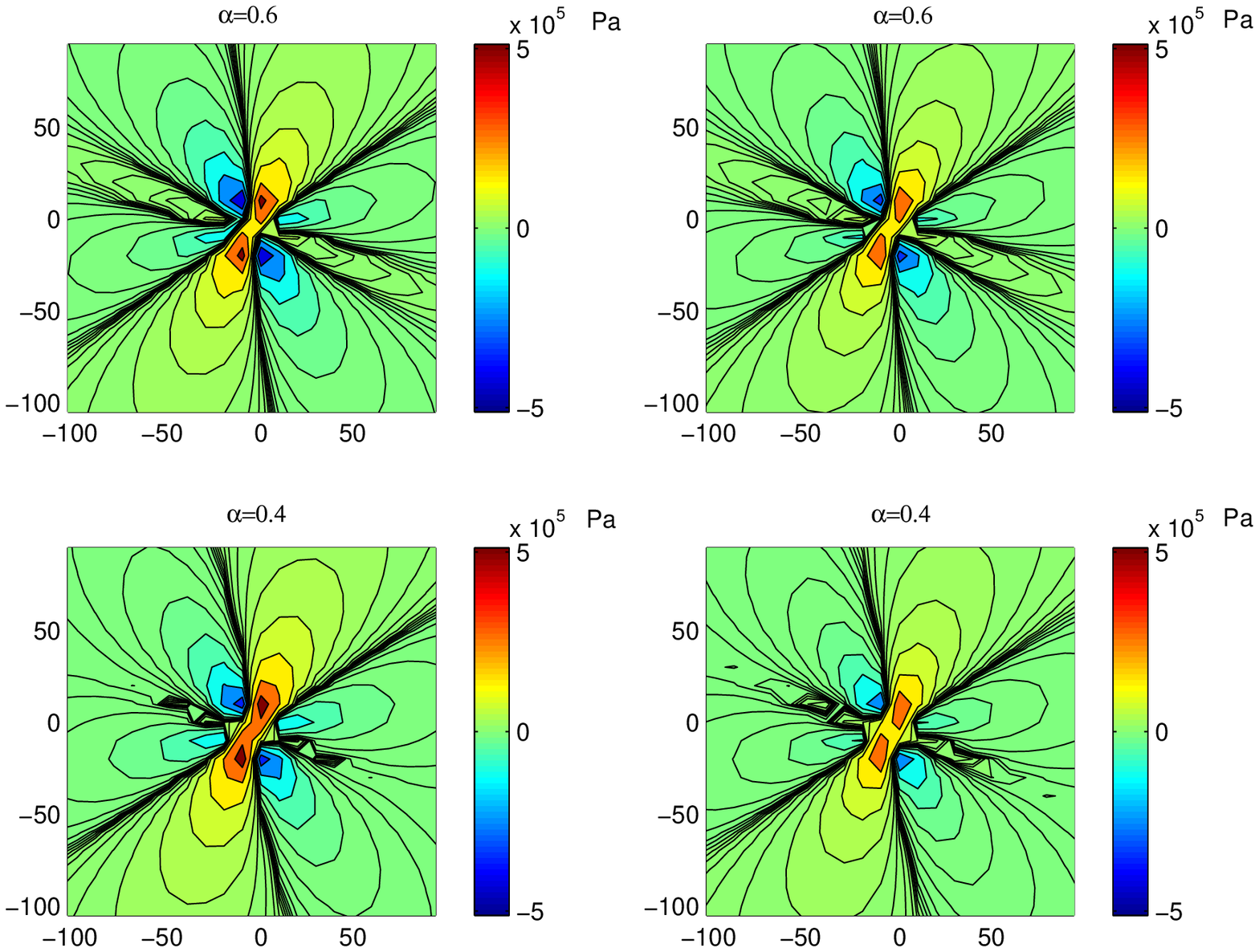, width=16cm}
\caption{\label{syy_60_40} Same as Fig.~\ref{syy_100_80}
for $\alpha=0.6$ and $\alpha=0.4$.}
\end{figure}

\begin{figure}
\epsfig{file=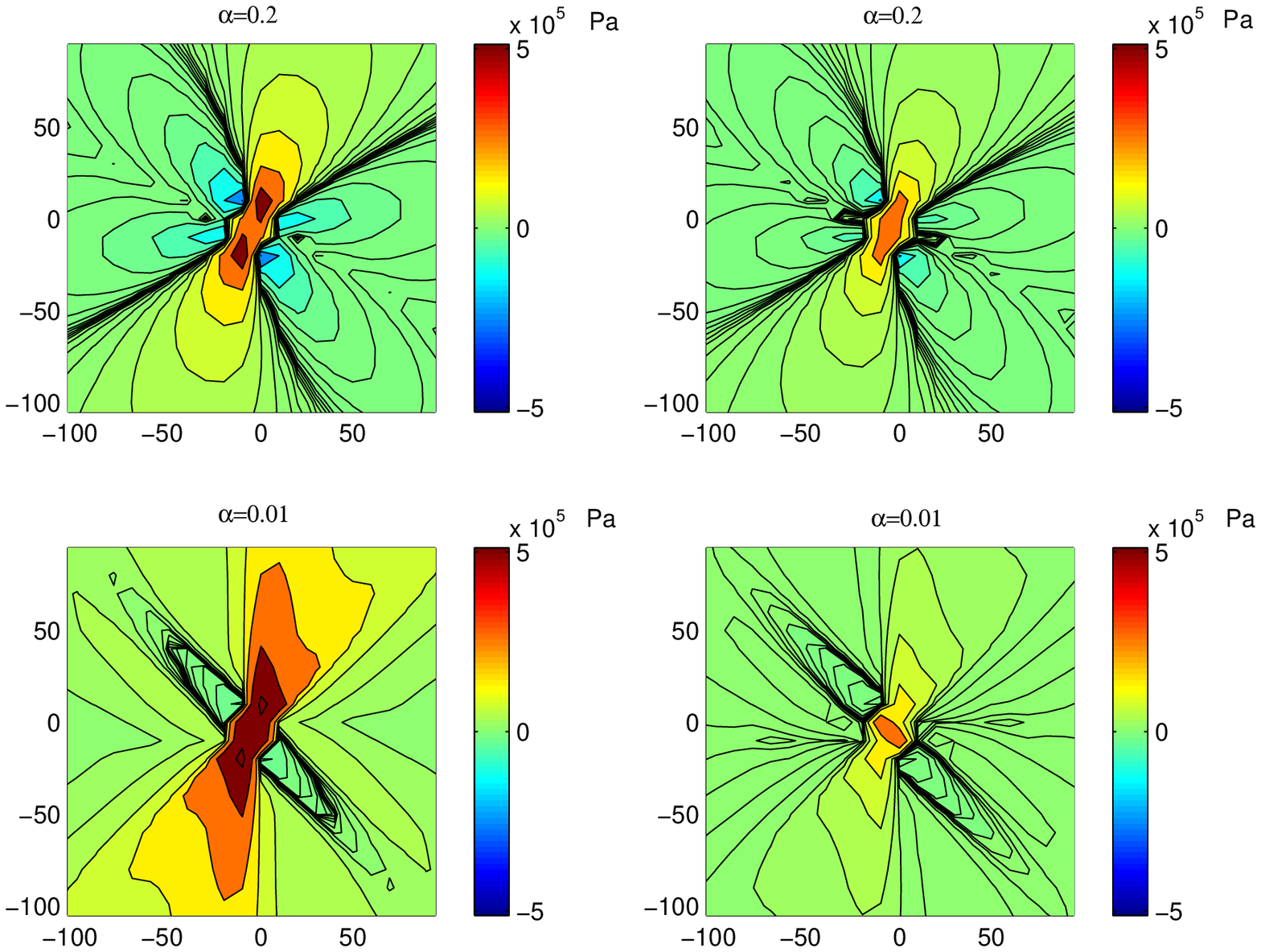, width=16cm}
\caption{\label{syy_20_1} Same as Fig.~\ref{syy_100_80}
for $\alpha=0.2$ and $\alpha=0.01$.}
\end{figure}

Figures \ref{sxx_100_80} to \ref{sxx_20_1} show the variation
of the stress component $\sigma_{xx}$ within the plate. The same kind of comments apply here 
as for $\sigma_{xy}$. This is also the case for component $\sigma_{yy}$ which is shown
in Figures \ref{syy_100_80} to \ref{syy_20_1}.

We have thus introduced a mechanical asymmetry at the microscopical level (i.e. at
the spring scale), for all springs'orientations corresponding to an isotropic asymmetry,
which translates into a loss of symmetry at the macroscopical scale. We shall quantify
this loss of symmetry more precisely in the next section.

\subsection{Decay of the stress amplitude away from the source}

If we consider the center of the source cell as the origin of our frame, every point
within the plate can be located using polar coordinates $(r,\theta)$, where $r$ is
the distance to the origin (the source cell center), 
and $\theta$ is an azimuth measured clockwise from the $\vec{O}_x$ axis. 
We can then, for a fixed $\theta$, look at the decay rate with $r$ of the modulus of any
of the stress tensor components. 
To avoid problems due to the finite size of the source and of the whole plate, 
we quantify the decay of the modulus of any stress component by a power-law of
the type $r^{-\gamma}$, within a distance interval bracketed within a few cell to a few
tens of cell sizes. We then repeat the same computation for different values of $\theta$. 
We then change the value of the asymmetry factor $\alpha$ and obtain the corresponding
dependences $\gamma (\theta)$ for each value of $\alpha$.

\begin{figure}
\epsfig{file=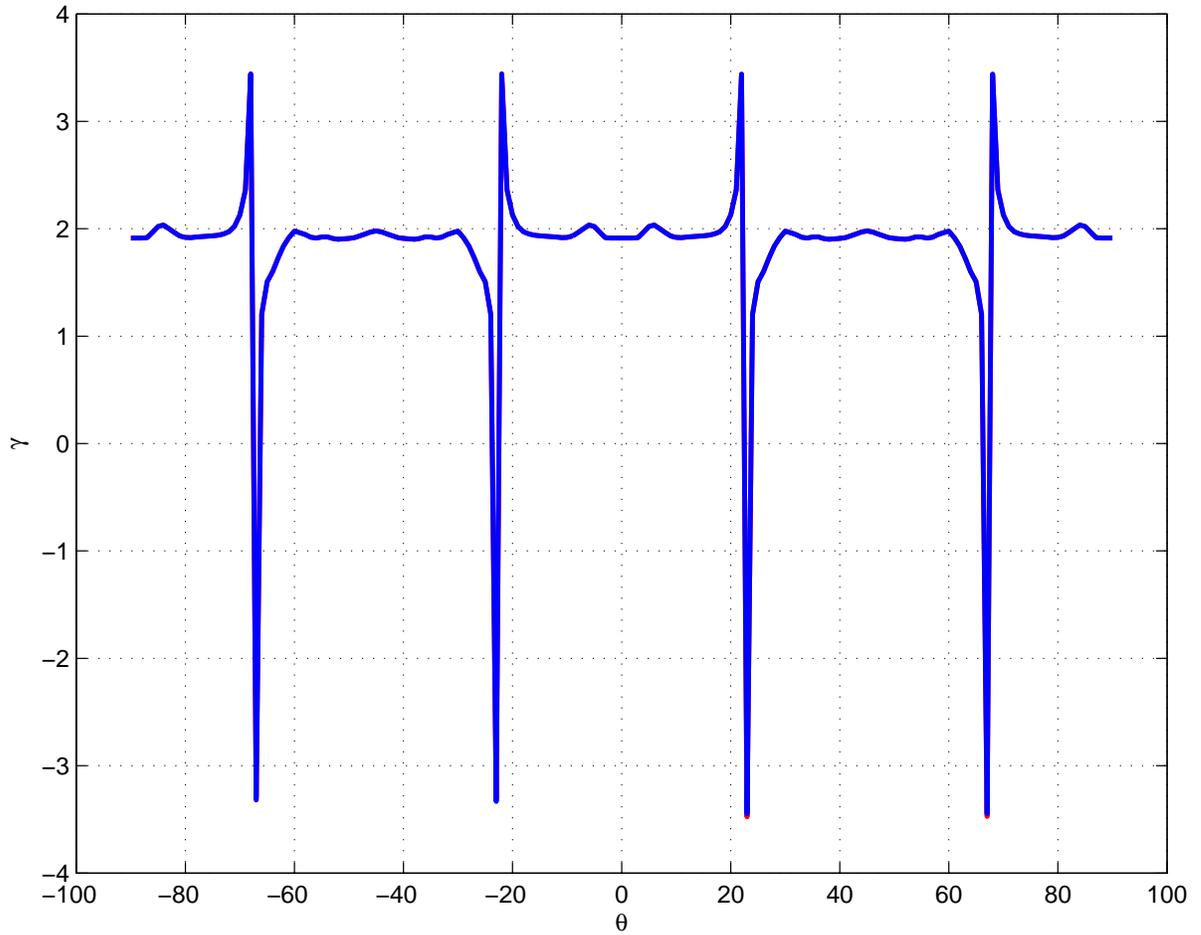, width=16cm}
\caption{\label{gamma_sxy_100} Variation of the decay exponent $\gamma$ with azimuth $\theta$ 
when $\alpha=1$ measured for the stress component $\sigma_{xy}$. The values of
$\gamma$ obtained with
pure shear stress loading and pure shear strain loading
superimpose exactly. All values of $\gamma$ are close to the theoretical value $\gamma=2$
for symmetric elasticity,
except at stress lobes borders where it is ill-defined numerically 
and exhibits spurious peak values.}
\end{figure}

\begin{figure}
\epsfig{file=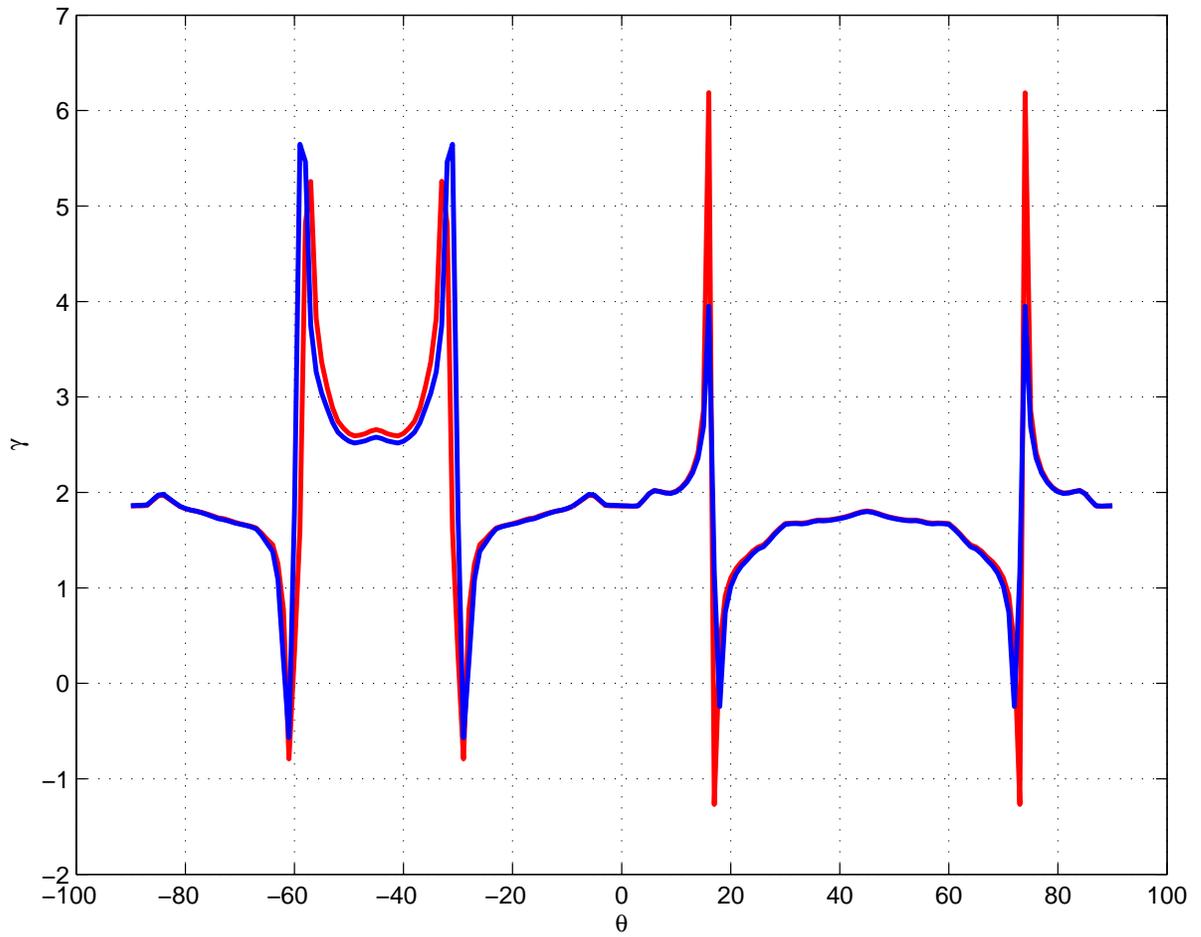, width=16cm}
\caption{\label{gamma_sxy_50} Variation of the decay exponent $\gamma$ with azimuth $\theta$ 
for $\alpha=0.5$ for the stress component $\sigma_{xy}$. 
The red (respectively blue) curve is for pure shear stress (respectively
pure shear strain) loading. Peak values are spurious and correspond to lobes boundaries.
The exponents are essentially identical for both types of loading conditions.}
\end{figure}

\begin{figure}
\epsfig{file=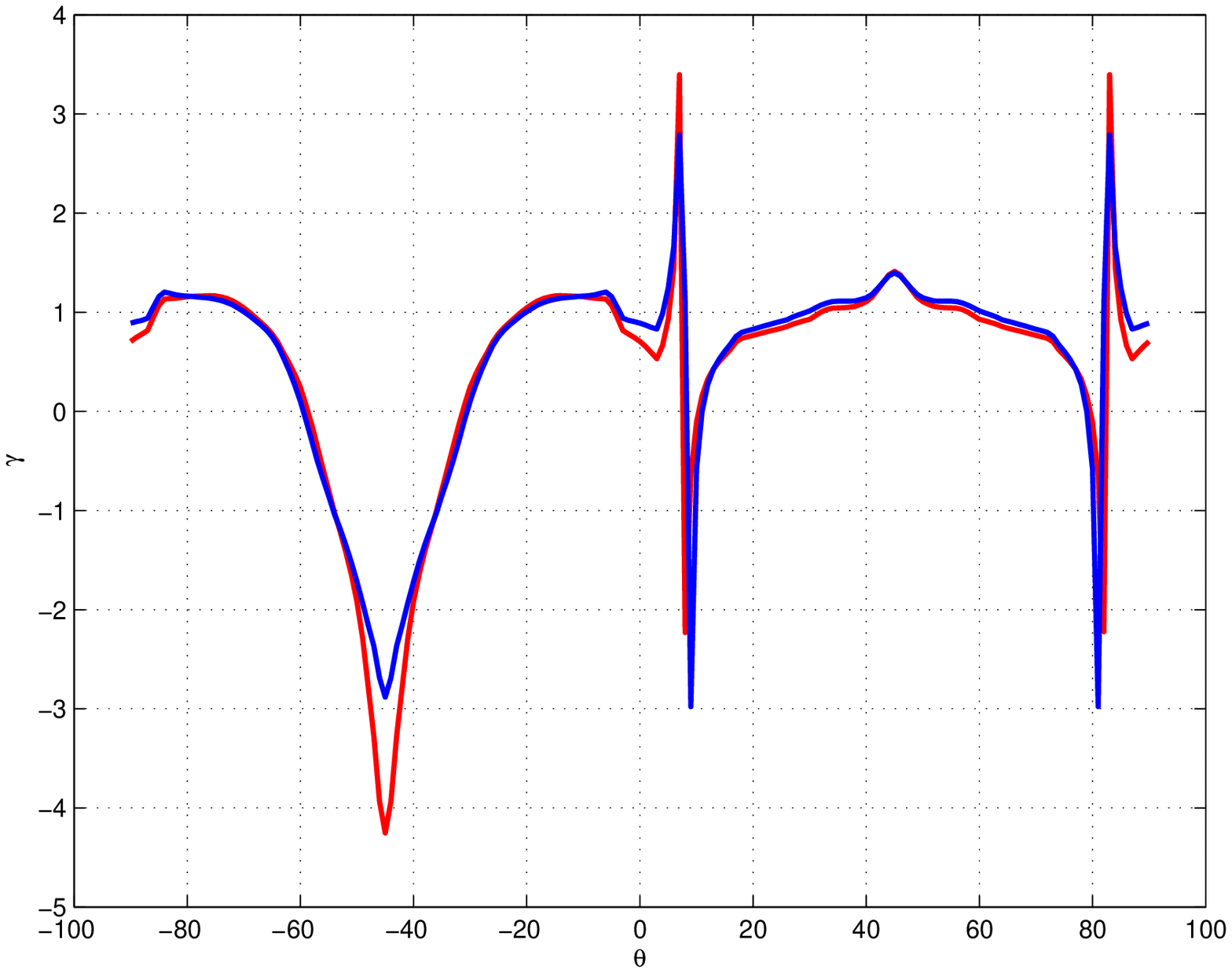, width=16cm}
\caption{\label{gamma_sxy_1} Same as Fig.~\ref{gamma_sxy_50}
for $\alpha=0.01$.}
\end{figure}

Figures \ref{gamma_sxy_100} to \ref{gamma_sxy_1} show the variation
of $\gamma$ with $\theta$, for different values of $\alpha$, quantifying the decay of the
$\sigma_{xy}$ component. Each frame features two
curves, one corresponding to the pure shear stress condition at the source (in red), the other
one to the pure shear strain condition (in blue).

When $\alpha=1$, $\gamma$ is very close to $2$, which is the theoretical value predicted
by planar elasticity. There are some small fluctuations around that value, the largest ones
being obtained for values of the azimuth corresponding to a change of sign of the stress, i.e. where
the stress itself almost vanishes. In those peculiar directions, the determination of the
exponent is very unstable.

When $\alpha$ decreases, $\gamma$ values can reach values very different from $2$. Some of those
values correspond to azimuths where stress vanishes (and are thus spurious), but others reflect
genuine consequences of the nonlinearity of the medium. One can see that exponents can thus reach
values larger than $2$ (reflecting a very rapid decay and thus
short interaction range), but that they can also get down
to values around or lower than $1$, leading to very large interaction ranges. When $\alpha=0.01$, 
one should not consider negative values of $\gamma$ too seriously as such negative
values would imply that the stress
increases with distance.  The increase is occurring only over a finite distance range
and gives way to a decrease at larger distance. The measured exponent is thus only 
valid at very short distances and is not an asymptotic value. We are unable for those cases
to quantify accurately the value of the asymptotic $\gamma$ due to finite size effects.
For that $\alpha$ values, one should thus consider that the asymptotic value of $\gamma$ varies 
en general between $0$ and $1$.

The found values of $\gamma$ as a function of $\theta$ also reveal 
that exponents do not vary significantly with the conditions imposed at the source, which 
constitutes
another surprise. However, stresses obtained in the pure shear strain condition are lower 
than in the pure shear stress condition, as
the prefactor of the power-law decay is found smaller than for the pure shear stress case.

\begin{figure}
\epsfig{file=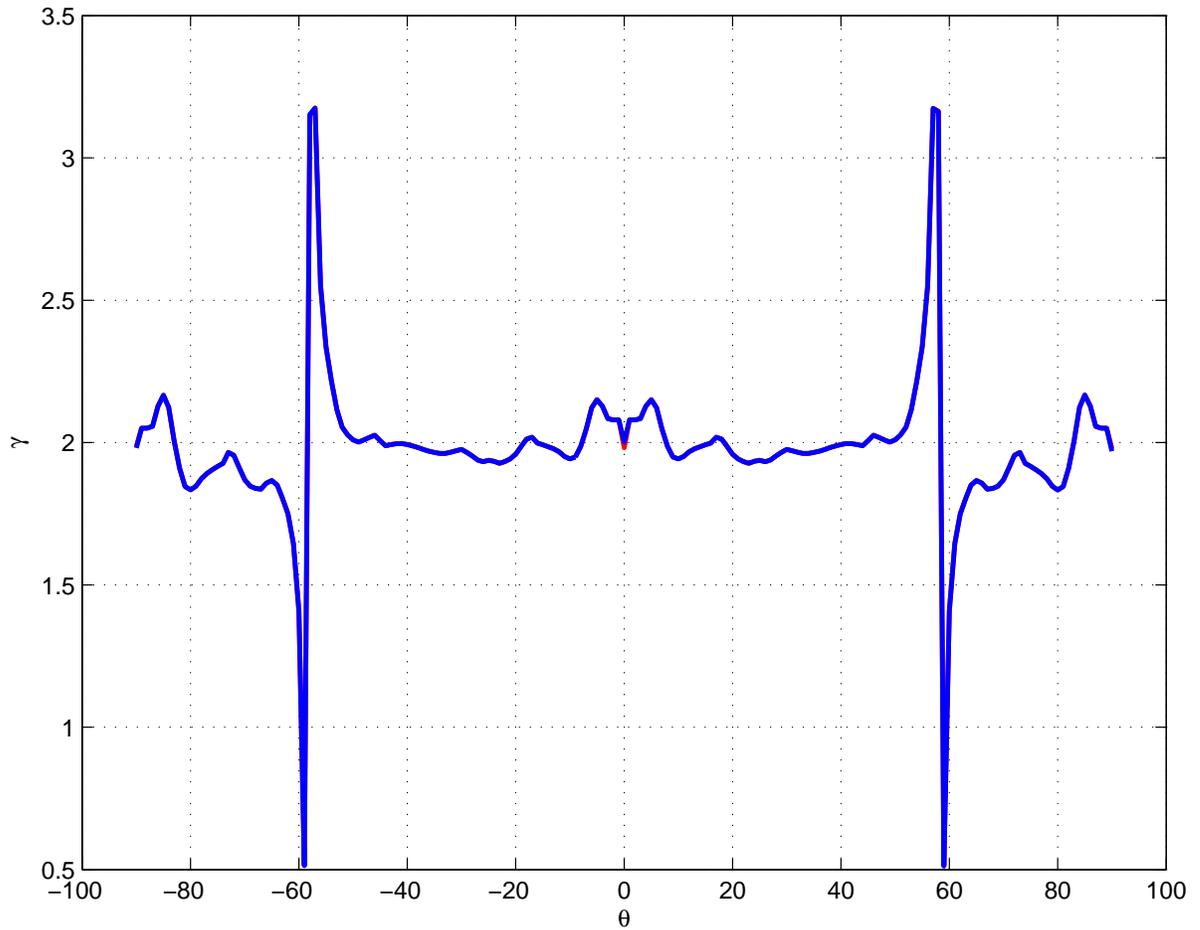, width=16cm}
\caption{\label{gamma_sxx_100} 
Variation of the decay exponent $\gamma$ with azimuth $\theta$ 
for $\alpha=1$ for the stress component $\sigma_{xx}$. 
The values of $\gamma$ obtained with pure shear stress and pure shear strain loading
superimpose exactly. $\gamma$ is found close to the theoretical value $2$,
except at stress lobe borders where it is ill-defined and exhibits spurious peak values.}
\end{figure}

\begin{figure}
\epsfig{file=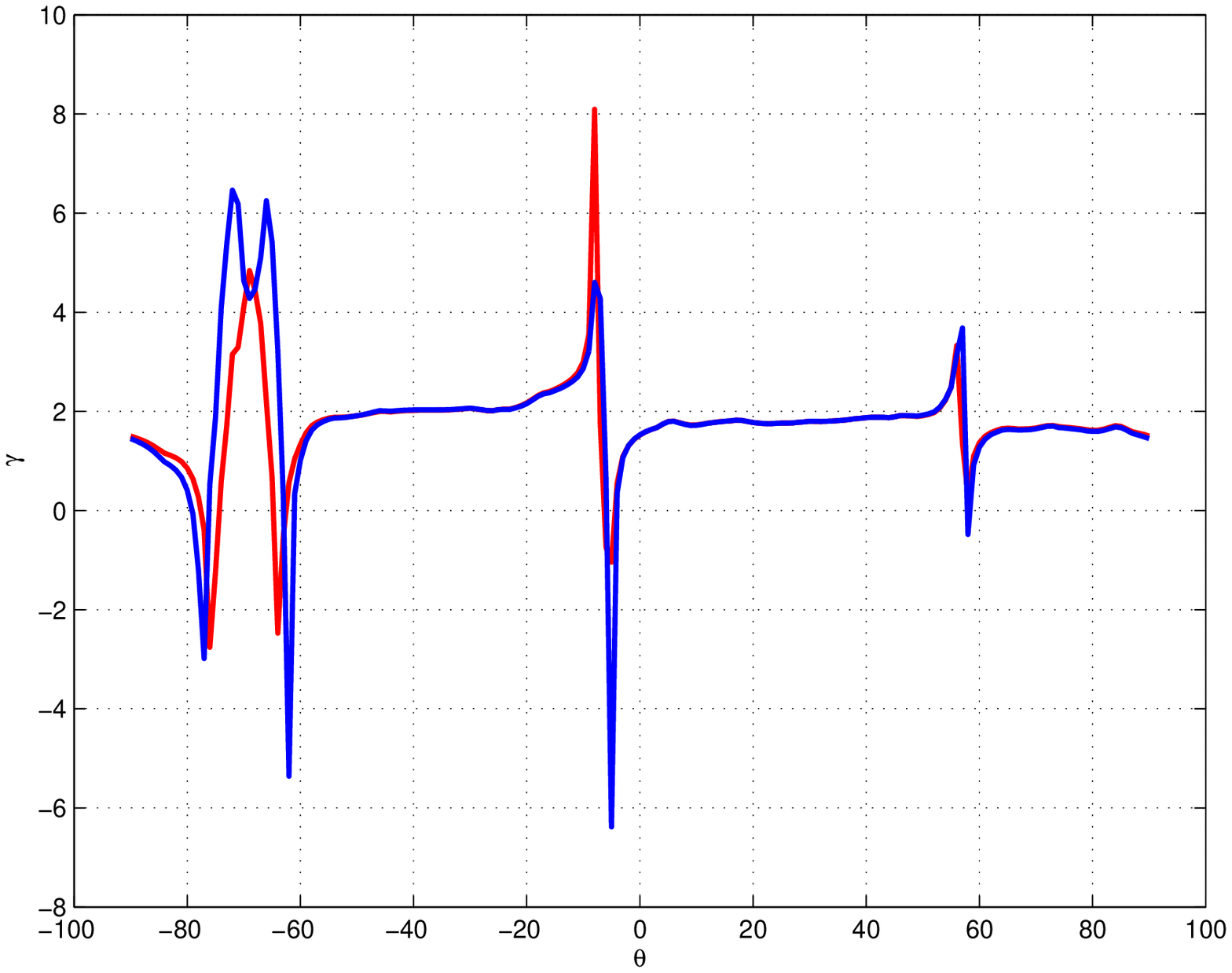, width=16cm}
\caption{\label{gamma_sxx_50} Same as Fig.~\ref{gamma_sxx_100}
for $\alpha = 0.5$.
The red (respectively blue) curve is for pure shear stress (respectively strain)
loading. Peak values are spurious and correspond to lobe boundaries.
The exponents are very similar for both types of loading conditions.}
\end{figure}

\begin{figure}
\epsfig{file=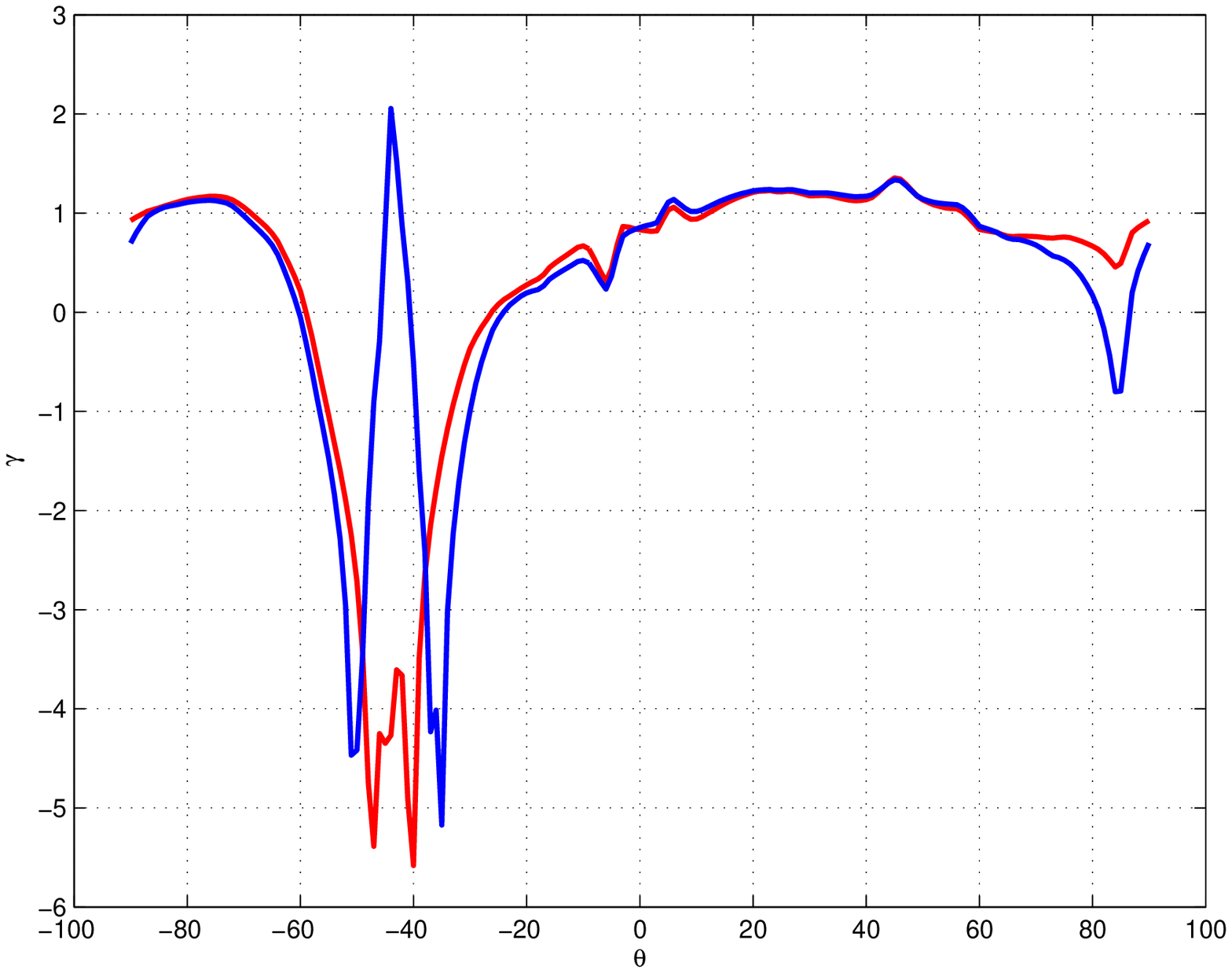, width=16cm}
\caption{\label{gamma_sxx_1} Same as Fig.~\ref{gamma_sxx_100}
for $\alpha = 0.01$.}
\end{figure}

\begin{figure}
\epsfig{file=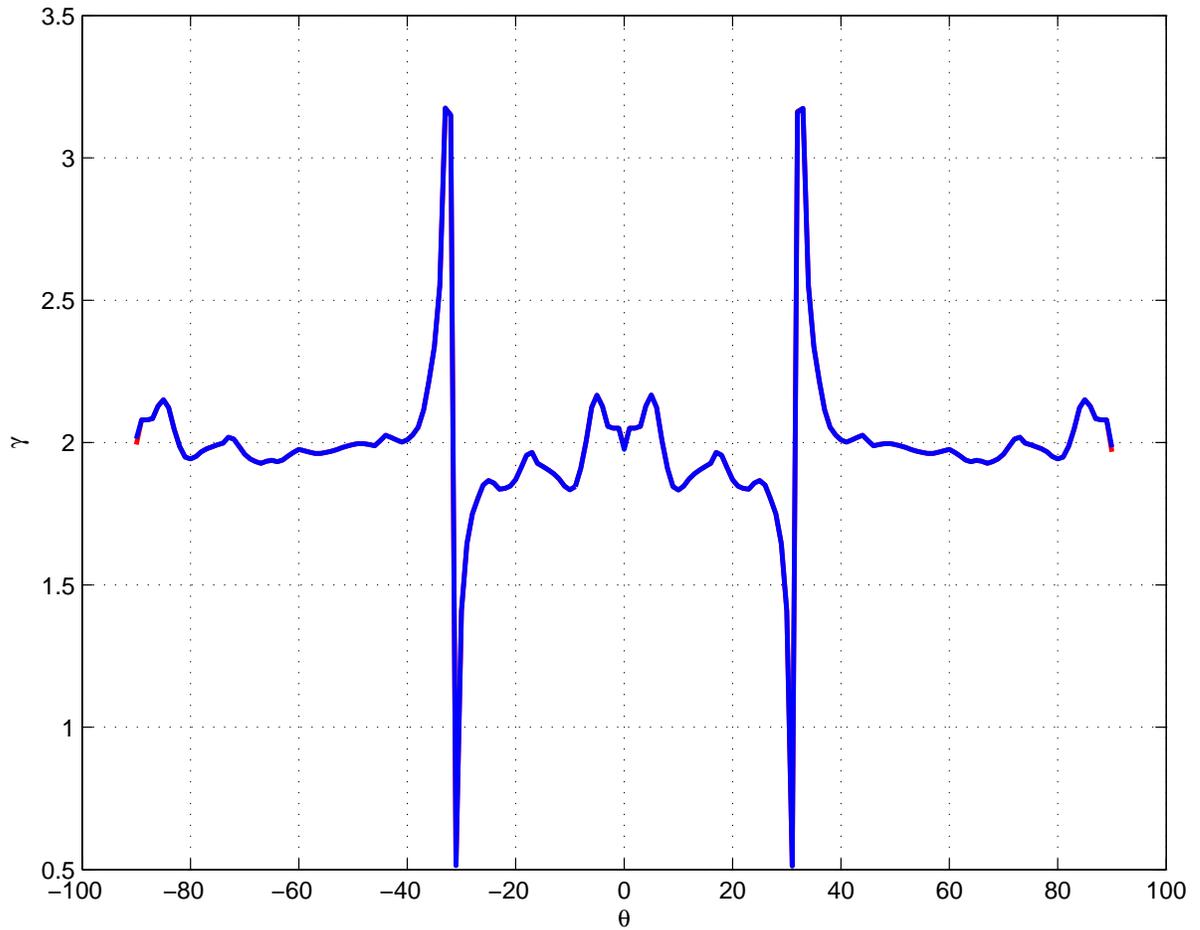, width=16cm}
\caption{\label{gamma_syy_100} Variation of the decay exponent $\gamma$ with azimuth $\theta$ 
for $\alpha=1$ for the stress component $\sigma_{yy}$. 
The values of $\gamma$ obtained with pure shear stress and pure shear strain loading
superimpose exactly. All values of $\gamma$ are close to the theoretical value $\gamma=2$,
except at stress lobe borders where they are ill-defined and exhibit spurious peak values.}
\end{figure}

\begin{figure}
\epsfig{file=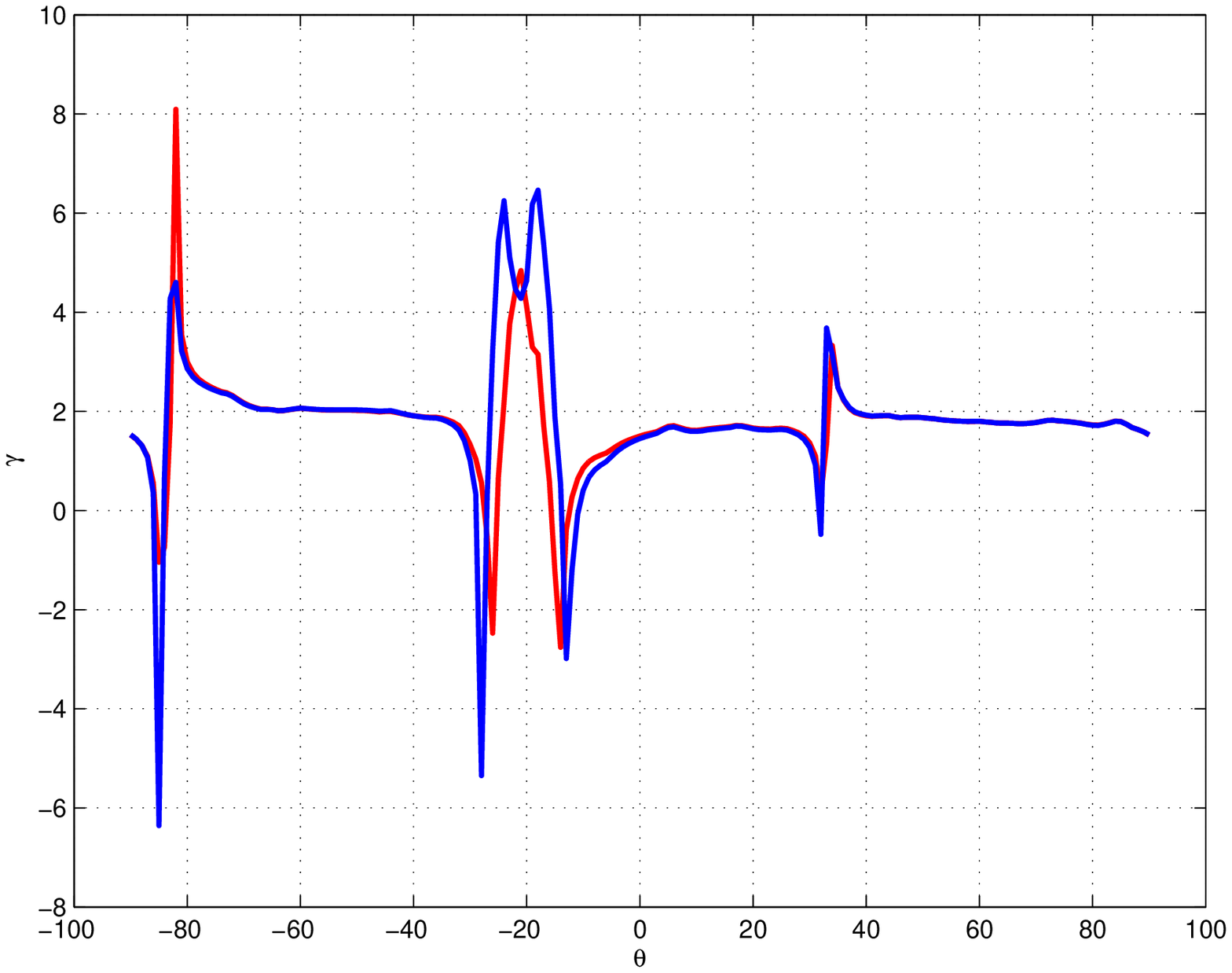, width=16cm}
\caption{\label{gamma_syy_50} Same as Fig.~\ref{gamma_syy_100}
for $\alpha=0.5$. 
The red (respectively blue) curve is for pure shear stress 
(respectively strain) loading. Peak values are again 
spurious and correspond to lobe boundaries.
The exponents are very similar for both types of loading conditions.}
\end{figure}

\begin{figure}
\epsfig{file=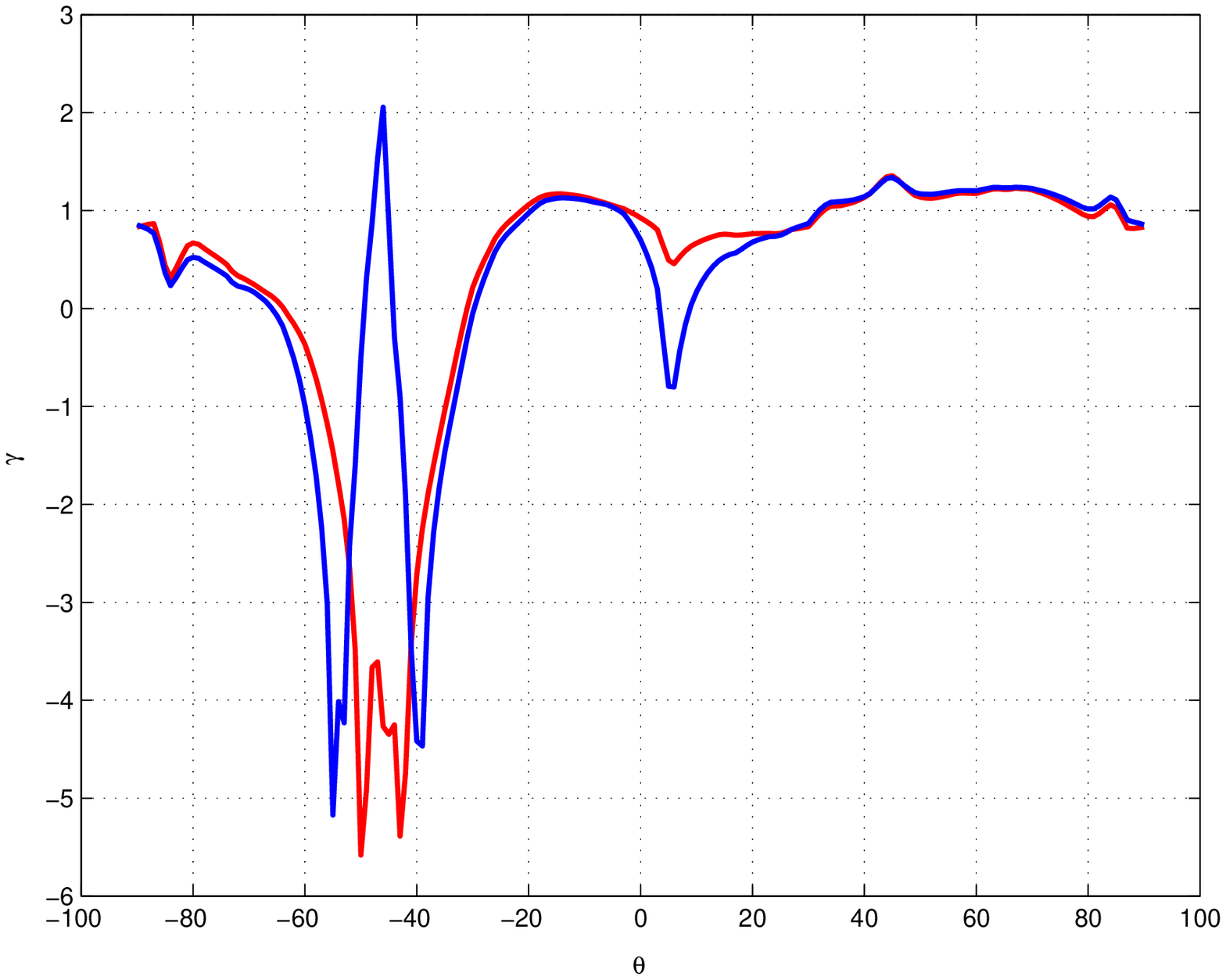, width=16cm}
\caption{\label{gamma_syy_1} Same as Fig.~\ref{gamma_syy_50}
for $\alpha=0.01$.}
\end{figure}

Figures \ref{gamma_sxx_100} to \ref{gamma_sxx_1}, as well as
Figures \ref{gamma_syy_100} to \ref{gamma_syy_1} show the same
results for components $\sigma_{xx}$ and $\sigma_{yy}$. For both components, and for
$\alpha=1$, we recover the theoretical value $\gamma=2$ for any $\theta$. As $\alpha$
decreases, the exponents can take very different values, including some which imply very
long range decay. We observe again that the exponents do not vary with the type of loading
at the source. We also checked that the explaination for negative values of exponents
was the same as for $\sigma_{xy}$. Thus, for low $\alpha$ values, $\gamma$ decreases to
values between $0$ and $1$.

\section{Discussion}

This idea of mechanical asymmetry, and/or of the feedback between local damage and stress decay
from perturbative sources is not new but, to the best of our knowledge,
 it is the first time that it is implemented in
a real $2D$ plane elastic problem applied to Earth mechanics and seismotectonics.
For example, \cite{roujks}
gives the analytical solution for the stress field and for the dependence $\gamma(\alpha)$ in 
a nonlinear asymmetric elastic medium in the case of antiplane mode III loading,
which is thus the scalar equivalent to the problem studied here. In the antiplane
case, there is only one stress component and a single exponent $\gamma(\alpha)$.
In this antiplane case, it can be shown analytically that the exponent $\gamma(\alpha)$
is indeed decreasing from the value $2$ for $\alpha =1$ to smaller values as 
$\alpha$ decreases. But there is not dependence on azimuth for this scalar case.

The existence of an asymmtry in the crust elasticity
has been proposed on the basis of observations of the Manyi 
($Mw=7.6$) earthquake \cite{Peltzer}. 
Using SAR interferometry data, Peltzer et al. \cite{Peltzer} interpreted
the mismatch between the displacement across each side of the 
left-lateral strike-slip fault
as due to a mechanical asymmetry between dilational and compressional quadrants.
Using a first-order perturbative calculation, they estimated a coefficient
$1/4 \leq \alpha \leq 1/2$ to explain the observed displacement asymmetry. However, they 
did not consider the possibility the
asymmetry could modify the long range decay rate of the stress field. This in turn
can modify 
the future seismic history in the neighborhood of this event. Their computation showed that
the asymmetric effect was probably confined in the very shallow part of the crust which, if
true, implies that the stress transfer at seismogenic depths after this event obeys standard 
linear elastic solutions.

The fact that this model should be relevant for shallow crustal mechanics is rather obvious
(as shown from the previous field example as well as from the short discussion at the beginning of 
this paper). The important question is to check if this model also holds (at least in limited
spatial domains) at depth. In that case, stress transfer case-studies should take account
of this asymmetry effect, which can greatly enhance the distance at which a given event can trigger
another one. Testing this hypothesis is not  simple, as the rheology we assumed is 
nonlinear, which means that the effect of successive events can not be simply added. 
Stress field evolutions
with time may then have much sharper transitions in space and time than predicted
by models involving linear elasticity,  
a behaviour reminiscent of the mechanics of granular media \cite{Bouchaud}. 
The consequence is that, to compare with the standard stress transfer 
mechanism \cite{Stein0}, we need
to know in details the state of stress within the crust prior to an event to map
predicted stress transfer lobes onto aftershocks location catalogs.

Another possibility for testing our hypothesis
would be to study the statistics of seismic moment tensors of events,
as we saw that, if we assume that the source can be describe dynamically as a pure double couple of
forces, this tensor should display a non-vanishing isotropic part. However, the seismic moment tensor
is computed from seismic wave observations (i.e. of dynamical nature), 
and not from static displacements in situ at depth.
Morover, the time at which the solutions we computed really hold depends on the diffusion properties
of fluids in rock. This is why such a way of testing would indeed imply to compute the whole
time-dependent dynamical asymmetric poro-elastic solution to really propose quantitative results
allowing a reliable comparison.

Other data that could be used to test the model are the SAR data, in the spirit of the work of
\cite{Peltzer}. Interpretation of SAR data done after a large event could prove the
pertinence of our model, provided that stress field evolution after the event is not modified
by other events or by the far-field loading of the plate.

Proving or refuting this model is of prime importance for the understanding of the spatio-temporal
patterns of earthquakes, including those preceeding a large and potentially destructive event.
We already discussed the fact that it could improve the potential of prediction methods based on
concepts such as stress transfer. But it has even deeper implications on the hope of predicting
large events from the behavior of the statistics of populations of shocks preceeding that event.
In another numerical work, Ref.~\cite{mora} studied the progressive damage of a fault plane
before its macroscopic rupture. Their model employs cellular automaton techniques to simulate tectonic 
loading, rupture events and strain redistribution. Note that in that case, strain is equivalent 
to stress. The elastodynamic Green function for stress/strain redistribution is taken to vary 
as $1/r^p$, where $p$ is a parameter which is varied. 
The systems displays two different regimes depending on the $p$ value.
For $p \leq 2$, large events are preceeded by a clear power-law acceleration of energy release 
of the system, together with the growth of strain energy correlations. This is of course
reminiscent of the critical earthquake hypothesis \cite{Bowetal,Bowmanetal,Sorsammis2}.
For $p$ larger than $2$, 
the trend of energy release before a large event is linear. This means that the lower $p$ is, the
more predictable is the large event, using time series of precursory
energy release. Their model does not map
exactly to ours,, but we could expect that if $\alpha$ is under a certain threshold (still to
be determined), then $\gamma$ would be low enough for the critical earthquake
scenario to apply, making large events predictable from time series. 
In the other hand, if in some areas $\alpha$ is
above that threshold, then the local tectonic domain would belong to the other regime, 
and large events would be unpredictable. If the predictability of large events relies, as suggested
by \cite{mora}, on the exponent of the Green function of stress transfer, then we have pointed out
a very simple physical mechanism allowing to tune that exponent. Large scale fluctuations of
fluid pressure from lithostatic to infra-lithostatic could then explain why in some cases large
events are preceeded by strain energy release acceleration, 
while the opposite holds in other cases.

\newpage

\clearpage

\clearpage

\clearpage

\clearpage

\clearpage

\clearpage

\clearpage

\clearpage

\clearpage

\clearpage

\clearpage

\clearpage

\clearpage

\clearpage

\clearpage

\clearpage

\clearpage

\clearpage

\clearpage

\clearpage

\clearpage

\end{document}